\documentstyle[12pt]{article}

\renewcommand{\H}{I\!\!H}
\newcommand{\R}{I\!\!R}

\begin{document}
\begin{titlepage}

\title{Decomposition and unitarity in quantum cosmology 
         \thanks{
         Work supported by the Austrian Academy of Sciences
         in the framework of the ''Austrian Programme for
         Advanced Research and Technology''.}}

\author{Franz Embacher\\
        Institut f\"ur Theoretische Physik\\
        Universit\"at Wien\\
        Boltzmanngasse 5\\
        A-1090 Wien\\
        \\
        E-mail: fe@pap.univie.ac.at\\
        \\
        UWThPh-1996-64\\
        gr-qc/9611006 
                      } 
\date{}

\maketitle

\begin{abstract}
Considering quantum cosmological minisuperspace models with positive
potential, we present evidence that
\par
{\it (i)} despite common belief there are perspectives for defining
a unique, naturally preferred decomposition of the space $\H$ of 
wave functions into two subspaces $\H^\pm$ that generalizes the concept 
of positive and negative frequency, and that 
\par
{\it (ii)} an underlying unitary evolution within these two 
subspaces exists and may be described in analogy to the representation 
of a geometric object in local coordinates: it is associated with the 
choice of a congruence of classical trajectories endowed with a 
suitable weight (such a setting is called WKB-branch).
\par
The transformation properties of various quantities under a
variation of the WKB-branch provide the tool for defining the
decomposition. The construction leads to formal series
whose actual convergence seems to require additional 
conditions on the model (related to 
global geometric issues and possibly to analyticity). 
It is speculated that this approach might relate to the refined 
algebraic quantization program. 
\medskip

PACS-number(s): 98.80.Hw, 04.62.+v$\,\,$.
\end{abstract}

\end{titlepage}

\section{Introduction}
\setcounter{equation}{0}

Minisuperspace quantum cosmology bears some formal similarity to
the quantization of a scalar particle on a curved space-time 
background. The Wheeler-DeWitt equation in the former corresponds
to the Klein-Gordon equation in the latter type of models. 
Therefore, concepts, methods and experiences that 
have originally emerged in particles quantization issues 
are often taken over to quantum cosmology. 
One of the most prominent statements that have a clear meaning in 
particle quantization is that in the generic case, i.e. if the
wave equation does not admit any symmetries, the is no
unique decomposition of wave functions into positive and negative
frequency subspaces. This statement has been transferred to quantum 
cosmology 
although in this field it is much less clear what positive and negative
frequencies are and what other concepts might serve as 
generalizations thereof. Also, it is often believed that the space of 
wave functions cannot --- in a natural way --- be decomposed into a direct 
sum such that the Klein-Gordon scalar product is positive definite on
one and negative definite on the other subspace. However, 
strictly speaking, this type of belief is true only
if one refers to the construction of a decomposition by means
of underlying local symmetries. 
\medskip

The purpose of this article is to show that 
(assuming the potential in the wave equation to be positive) 
there are still perspectives in a number of models 
for the construction of a preferred decomposition generalizing 
the concept of positive and negative frequency. 
In any model admitting the application of the procedure
proposed here, 
this decomposition is not based on a local symmetry 
but rather on global issues. It gives rise to a Hilbert space 
structure at the fundamental level of solutions to the
(minisuperspace) Wheeler-DeWitt equation. 
Furthermore, in situations where such a preferred decomposition 
exists, a unitary evolution within the two
subspaces is exhibited in a local context, 
depending on the choice of a classical action $S$ and a 
weight function $D$. Calling the pair $(S,D)$ a ''WKB-branch'', 
the hermitean evolution operator emerges in some analogy to the
way tensor components appear in coordinate systems, namely as
representatives of some mathematical ''object''. 
The proposal we formulate is understood as a general strategy whose
realization in concrete models will face specific problems 
that might be easy or hard (or even impossible) to solve. 
\medskip

A further goal of this article is to speculate on the relation 
between the preferred decomposition and structures that appear
in the context of other quantization methods, in particular
the refined algebraic quantization. This method incorporates 
global issues from the outset and ends up with a positive 
definite inner product on the space of physical states. 
Uncovering a relation to the more conventional quantization 
in terms of the Klein-Gordon scalar product could eventually 
lead to a more fundamental and unified understanding of what happens 
when one ''quantizes'' a theory. 
\medskip

The paper is organized as follows: In Section 2, we 
write down the wave equation and the two scalar products 
(Klein-Gordon and $L^2$) at hand. Also, we briefly 
review how congruences of classical trajectories in 
minisuperspace are described by the Hamilton-Jacobi 
equation and its solutions. 
After a discussion of the decomposition issue and the 
refined algebraic quantization for the case of the 
Klein Gordon equation in flat space-time, we formulate the 
general problem. 
\medskip

Section 3 is devoted to the detailed 
presentation of 
a formalism serving to analyze the wave equation in a novel way. 
Any WKB-branch may be used as a ''background structure'' with 
respect to which many decompositions are possible, each one
associated with a solution of a differential equation for
an operator called $H$. 
This differential equation basically states that 
in each of the two subspaces the wave equation is of the 
Schr{\"o}dinger equation type, $H$ 
serving as evolution operator. 
The key idea of this procedure is to generalize the 
notion of outgoing and incoming orientation of {\it classical} 
trajectories to the {\it quantum} case. At first sight, 
this construction is not unique, 
and we end up with a variety of ways to do so. 
\medskip

In Section 4, we analyze the transformation properties of 
various quantities when the WKB-branch used as ''background
structure'' is {\it changed} infinitesimally. 
Section 5 is devoted to the formulation of our proposal. 
At a formal level we show that --- within each
WKB-branch --- one may choose a solution 
$H$ of the above-mentioned differential equation 
such that {\it all decompositions} emerging
in different WKB-branches {\it agree with one another}. 
This opens the perspective 
to unambigously single out {\it one preferred decomposition} 
of the space of wave functions. 
The question whether the formal (power series) solution for $H$ 
actually convergences (or may consistently be regularized) 
is addressed to the concrete models upon 
which the strategy shall be applied. 
We explicitly display the first few terms
in the expansion of $H$ (and some redefined version $\cal K$ 
which makes explicit the unitarity of the evolution). 
In semiclassical contexts, these terms may be expected
to approximate the actual solution and should thus unambigously
describe the decomposition and the underlying unitarity.  
\medskip

In Section 6, we comment on the structure exhibited 
by the formal solution of the decomposition problem 
and give some speculations on its nature. It is clear that
--- whenever it exists --- the solution 
refers to the {\it global structure} 
of minisuperspace (possibly by means of some average over
all possible decompositions), and there are hints that some 
{\it analyticity} structure might play a role as well. 
In Section 7, we display simple examples for which
the operator $H$ may be expressed in terms of known
functions. 
Furthermore, using a toy-model situation, we 
show that the obvious heuristic argument against a preferred
decomposition (the difference of asymptotic incoming and 
outgoing decompositions on a time-dependent background) 
does not necessarily apply. 
Section 8 is devoted to 
speculations about relations of our porposal 
to the refined algebraic quantization scheme. 
We give a heuristic argument 
(whose basic ingredients date back to the literature of the 
Sixties) 
that the 
mere coexistence of the positive definite 
physical inner product constructed in this 
approach with the indefinite Klein-Gordon scalar product 
on the space of wave functions 
singles out a preferred decomposition. It is however not
clear whether it coincides with the decomposition emerging
from our proposal or whether it is yet another ''natural'' 
solution of the decomposition problem.  
Finally, in Section 9, we comment on possible generalizations 
to the case of
not necessarily positive potential and point out that 
this issue might touch upon
the significance of analytic continuation and the role of the
so-called Euclidean domains in minisuperspace. 
\medskip

Summarizing, our approach intends to clarify the relation between 
structures emerging in different quantization schemes. 
At a formal level, we can offer a procedure that starts from the
conventional setting of the Klein-Gordon scalar product on the
space solutions to the wave equation but transcends this framework 
and exhibits a Hilbert space structure at the fundamental 
level, without making use of semiclassical 
techniques and other approximations. 
Although constructed in a local context, this structure 
is likely to be based on global issues, and 
it might be related (or even identical) 
to the physical Hilbert space 
as constructed in the more advanced refined algebraic 
quantization method. 
\medskip

In order to improve the readability of this article, we list the
most important notations used:
\begin{tabbing}
symbol\qquad\qquad \= explanation \kill
$(S,D)$ \> Classical action and wave function prefactor, defined in
         (\ref{2.5})\\ \> and (\ref{3.6}). The pair $(S,D)$ is called
              WKB-branch.\\
$t$ \> $-S$, classical evolution parameter along trajectories, 
       serving\\ \> as a coordinate in minisuperspace $\cal M$.\\
$Q$ \> Indefinite Klein-Gordon type scalar product defined in 
       (\ref{2.8}).\\
$\langle\,|\,\rangle$ \> Positive definite scalar product, 
    defined by (\ref{3.13}) in terms of\\ \> a WKB-branch.\\
$\H$ \> Suitably defined vector space of solutions to the wave 
        equation \\ \> (\ref{2.2}), such that $Q$ exists on $\H$.\\
$\H^\pm$ \> Subspaces of $\H$, generalizing the notion of 
            positive/negative\\ \>  frequency.\\
$h$ \> Operator characterizing the wave equation in the form
       (\ref{3.9}), \\ \> defined in (\ref{3.10})--(\ref{3.11}).\\
$H$ \> Operator in the evolution equation (\ref{3.23})
       for wave functions\\ \> in $\H^+$, subject to the system of
       equations (\ref{3.18})--(\ref{3.20}).\\
$\cal K$ \> Hermitean evolution operator in the Schr{\"o}dinger
   type equation\\ \> (\ref{3.38}) for wave functions in $\H^+$, 
   defined in (\ref{3.40}).\\
$\epsilon$ \> Formal book-keeping parameter which can be inserted into 
           most \\ \> equations of this paper; its actual value 
           is $1$, and it is introduced\\ \> in Section 5.\\
${\cal H}_{{}_{\rm phys}}$ \> Physical Hilbert space as constructed
            by the refined algebraic \\ \> quantization method.\\
$\langle\,|\,\rangle_{{}_{\rm phys}}$ \> Positive definite inner 
            product as constructed by the refined \\ \> 
            algebraic quantization method.\\
\end{tabbing}
\medskip

\section{Klein-Gordon type wave equation with positive potential 
--- the problem}
\setcounter{equation}{0}

The basic mathematical ingredients for quantum cosmological
minisuperspace models 
\cite{Halliwell3}
as well as for scalar particle
quantization in a background space-time are an $n$-dimensional 
differentiable manifold $\cal M$ together with a metric $ds^2$ 
of pseudo-Riemannian signature $(-,+,+,\dots,+)$ and a 
real-valued function $U$ on $\cal M$. 
In many practical cases, these objects will be 
$C^\infty$ or even (real) analytic, and we postpone further
specification of their differentiability properties to 
speculations in Sections 5 and 6. 
In local coordinates $y\equiv (y^\alpha)$ we can write
\begin{equation}
ds^2 =g_{\alpha\beta}(y)\,dy^\alpha dy^\beta\, . 
\label{2.1}
\end{equation}
In quantum cosmological 
models, $\cal M$ (the minisuperspace) consists of the spatial 
degrees of freedom 
of geometry and matter, $ds^2$ is the DeWitt metric and $U$ the 
potential, stemming from the spatial curvature as well as the
matter couplings. In the scenarios of scalar particle 
quantization in a
background, $\cal M$ is the space-time manifold and $ds^2$
its metric. The massive (Klein-Gordon) case is given by constant
potential $U=m^2$, but more general potentials are admissible 
as well. 
\medskip

We require in addition that there exists a foliation of $\cal M$
by spacelike hypersurfaces. (The words ''spacelike'' and 
''timelike'' shall refer to the ''causal structure'' provided by
the metric $ds^2$. In quantum cosmology, this is of 
course {\it not} the causal structure of space-time).
As will be pointed out below in this Section, it will be 
necessary in some models to impose asymptotic conditions, 
e.g. to choose a foliation consisting of Cauchy hypersurfaces. 
In any case, 
we can assign each local light cone a future and a past 
orientation in a globally consistent way. 
Any smooth oriented timelike curve in $\cal M$  is 
either future- or past-directed. In the case of 
quantum cosmology, one could call these two orientations 
''outgoing'' and ''incoming'', respectively. 
The hypersurfaces providing the foliation can be labelled as 
$\Gamma_\lambda$ ($\lambda\in\R$), with future (outgoing)
orientation being equivalent to (or, likewise, being defined by) 
increasing $\lambda$, past (incoming) orientation by 
decreasing $\lambda$. 
\medskip

As an important restriction for what follows we require that the
potential $U$ is {\it everywhere positive} on $\cal M$. 
In quantum cosmology this implies that any classical
trajectory (representing a complete 
space-time with matter configuration) is everywhere timelike, 
i.e. either always outgoing
or always incoming. Returning trajectories and turning points
are excluded, as we shall see shortly. For a particle in a
background this condition guarantees that the classical motion 
(i.e. a world-line in the space-time manifold $\cal M$) respects 
the causal structure provided by the metric in that the four-velocity
is always timelike (for generality, we admit both future- and
past-oriented trajectories). In the massive 
case $U=m^2$, this condition is obviously satisfied for real $m$. 
In Section 9 we will speculate about the perspectives to relax 
the condition $U>0$. 
\medskip

Denoting by $\nabla_\alpha$ the covariant derivative with respect to
the metric, the basic wave equation is given by
\begin{equation}
\Big(-\nabla_\alpha \nabla^\alpha + U \Big)\,\psi = 0\, , 
\label{2.2}
\end{equation}
its solutions $\psi\equiv\psi(y)$ being denoted as wave functions, 
or ''wave functions of the universe''. They may be considered
as representations of the quantum states of the system.  
In quantum cosmology (\ref{2.2}) is the (minisuperspace version
of the) {\it Wheeler-DeWitt equation}, whereas in the particle
quantization models it may be called the {\it Klein-Gordon equation
in a background space-time with external potential}. 
\medskip

We should now make more explicit what we consider to be the 
underlying classical dynamics. A classical evolution of the system under 
consideration is provided by an oriented trajectory $y(t)$ in 
$\cal M$ satisfying the equations of motion generated by
the action 
\begin{equation}
S=\frac{1}{2}\int dt\, \Big(\,\frac{1}{N}\,g_{\alpha\beta}\, 
\dot{y}^\alpha\dot{y}^\beta - N U \Big).
\label{2.3}
\end{equation}
The ''lapse function'' $N$ serves as a Lagrangian multiplier and
thus does not have a dynamical evolution equation. Its appearance 
in the action expresses reparametrization invariance, 
i.e. it reflects the fact that the evolution parameter $t$ has
no physical significance (except for its orientation, which is achieved
by restricting $N$ to be positive). 
{\it After} variation with respect to the variables $(N,y^\alpha)$,
it may be chosen as a function $N\equiv N(y)$, thereby
fixing the evolution parameter $t$. We will not write down the
equations of motion explicitly but just note that the
momenta $p_\alpha$ are related to the ''velocities'' by 
\begin{equation}
\frac{1}{N} \, \frac{dy^\alpha}{dt} = g^{\alpha\beta}p_\beta \,, 
\label{2.4}
\end{equation}
and variation of $S$ with respect to $N$ yields the constraint 
$p_\alpha p^\alpha=-\,U$, whose quantum version is the wave equation
(\ref{2.2}).
Sometimes, for $n\ge 3$, when passing from the classical to the quantum
theory, the original classical potential $U^{\rm cl}$ is modified by 
adding a curvature term,  
$U=U^{\rm cl}+\xi R[g]$, where $\xi=\frac{1}{4}(n-2)/(n-1)$ and 
$R[g]$ is
the Ricci-scalar of the DeWitt metric \cite{Halliwell3}. This makes 
the wave equation invariant under
redefinitions of the lapse 
$\overline{N}=\Omega^2(y) N$, which amounts to a conformal
transformation 
$\overline{g}_{\alpha\beta}=\Omega^2 g_{\alpha\beta}$, 
a rescaling of the classical potential 
$\overline{U}^{\rm cl}=\Omega^{-2}U^{\rm cl}$ and a rescaling of
the wave function $\overline{\psi}=\Omega^{(n-2)/2}\psi$. 
In such a case the quantity $U$ in (\ref{2.3}) shall denote 
the modified potential, as it
appears in the wave equation, and when talking about
classical trajectories, we understand the dynamics generated
by the action (\ref{2.3}) with precisely the same potential $U$. 
In other words, classical and quantum theory are both
formulated in terms of the same fundamental objects $ds^2$ and $U$, 
the latter everywhere positive. In some cases 
with not everywhere positive potential, it is possible to 
achieve $U=U^{\rm cl}+\xi R[g]>0$
after a transformation of the type given
above. When trying to find an appropriate conformal factor $\Omega^2$,
one has to make use of the conformal transformation properties 
of the Ricci-scalar. One should however agree on a definite choice of 
$U$ to work with, because the formalism to be developed in this 
paper is not conformally invariant in an obvious way. 
(This would be an interesting issue to pursue, but we will not 
raise it here). 
\medskip

In what follows we will not make use of single classical trajectories
but of congruences generated by solutions of the Hamilton-Jacobi
equation. The latter reads
\begin{equation}
S_\alpha S^\alpha = -\, U\, , 
\label{2.5}
\end{equation}
whereby we use the abbreviation
$S_\alpha\equiv \nabla_\alpha S$, 
$S^\alpha\equiv \nabla^\alpha S$. 
Once having chosen a real solution $S(y)$ (which in general 
may exist only in a region $\cal G$ of $\cal M$), one may
identify its gradient $S_\alpha$ with the momentum
$p_\alpha$ of a family of oriented curves. Having fixed
$N$ to be some positive function on $\cal M$, 
(\ref{2.4}) plays the role of the differential equation 
\begin{equation}
\frac{1}{N} \, \frac{dy^\alpha}{dt} = S^\alpha 
\label{2.6}
\end{equation}
for $y^\alpha(t)$, the general solution being a congruence
(i.e. a non-intersecting $(n-1)$-parameter family) of 
solutions to the classical equations of motion. Redefining
$N$ just rescales the evolution parameter $t$ along the
trajectories (in an orientation-preserving way). Choosing a 
particular function $N$ eliminates the reparametrization 
invariance, and thus has the status of fixing a gauge. 
A Hamilton-Jacobi function $S$ is 
sometimes called ''classical action''. 
This is because $S$ agrees with (\ref{2.3}) when the integration is 
carried out along the according trajectories 
(and all lower integration 
bounds lie on a fixed hypersurface of constant $S$). 
\medskip

The tangent vectors of all trajectories
in the congruence are, by construction, orthogonal to the 
hypersurfaces $S=const$. Due to the positivity of the potential,
the trajectories are timelike, the hypersurfaces of constant
$S$ are spacelike. The trajectories belonging to some
$S$ are either {\it all} outgoing or {\it all} incoming. 
Thus, any $S$ may be associated with one of these two possible classical
modes of orientation. Accordingly,
$S'=-S$ generates the identical congruence, but with all trajectories
reversed in orientation. 
Whenever different trajectories intersect each other 
(e.g. on a caustic), or whenever the congruence contains 
a sequence approaching a lightlike curve, $S$ develops a singularity. 
(An example of this latter type is provided by 
$ds^2=-d\tau^2+dx^2$, $U=1$ and $S=\sqrt{\tau^2-x^2}$). 
This causes Hamilton-Jacobi functions $S$ in general to be
defined only in some region $\cal G$ of $\cal M$ 
(although the analytic continuation of $S$ 
across the boundary of $\cal G$ plays a role in certain approaches
to quantum cosmology).  
\medskip

After having summarized how the classical evolution emerges from
the structure provided by $ds^2$ and $U$, we return to the
issue of the quantum problem. There are two natural candidates
for a scalar product, namely
\begin{equation}
q(\psi_1,\psi_2)=\int_{\cal M} d\mu\, \psi_1^*\psi_2 
\label{2.7}
\end{equation}
and
\begin{equation}
Q(\psi_1,\psi_2)= -\,\frac{i}{2}\, \int_\Sigma d\Sigma^\alpha\,
(\psi_1^*\stackrel{\leftrightarrow}{\nabla}_\alpha \psi_2) \, , 
\label{2.8}
\end{equation}
where $d\mu \equiv d^n y \,\sqrt{-g}$ is the covariant volume 
element on $\cal M$, $\Sigma$ a spacelike hypersurface and 
$d\Sigma^\alpha$ the covariant hypersurface element on $\Sigma$. 
Both scalar products have been considered in attempts to 
quantize the system and to 
interpret
solutions of the Wheeler-DeWitt equation in terms of
observational physics. 
The first one, $q$, is sometimes modified into an integration
over a bounded domain of $\cal M$ and in this form intended 
to be the basis for computing relative (conditional) probabilities
for ''finding'' the configuration variables in the respective
domain. It is associated with 
what is sometimes called ''naive interpretation'' of quantum 
cosmology and has been
advocated by Hawking and Page \cite{HawkingPage}.
It is also the starting point for the ''refined algebraic
quantization'' program 
that is used in the more recent versions
of the Ashtekar approach \cite{Aetal} and about which we will
say a few words below. 
The second (Klein-Gordon type) indefinite expression $Q$ is 
especially designed 
for the case that $\psi_1$ and $\psi_2$ satisfy the
wave equation (\ref{2.2}) (in which case its
integrand is a conserved current) and is the starting point 
for another range of approaches to quantum cosmology, such as the 
Bohm interpretation \cite{Bohmetal} and the semiclassical (WKB) 
interpretation as proposed, among others, by Vilenkin 
\cite{Vilenkininter}. By replacing 
$\psi_1^*\rightarrow \psi_1$ and omitting the
prefactor $i$ in (\ref{2.8}), it is sometimes viewed as a
symplectic structure on the space of real wave functions. 
Some results have been achieved on the relation 
between the two scalar products $q$ and $Q$ 
\cite{HawkingPage,Higuchi2,Marolf1}, 
but they either involve approximations or refer to very
specifiy situation and are 
thus not very conclusive at the fundamental level. 
\medskip

The formalism to be developed in this article 
starts from the ''conservative'' point of view that some
appropriately defined space $\H$ of solutions of the wave
equation (\ref{2.2}) is endowed with the structure $Q$, and that
the physical significance of a wave function should be
expressed in terms thereof. One could call this point of view 
''Klein-Gordon quantization'' (cf. Ref. \cite{Kuchar1}). 
When proceeding, non-local and non-causal issues will appear, 
and we will frequently speculate on relations of our formalism
to schemes that are based on $q$, in particular the
refined algebraic quantization that was mentioned already 
above. This method has been developed during the past few
years by several authors 
(see Refs. 
\cite{Aetal,Higuchi2,Marolf1,Higuchi1,Marolf2}; 
also see Refs. \cite{Nachtmann,RumpfUrbantke,Rumpf2,Rumpf1}  
for predecessors).
In our simple case, the idea is essentially to utilize the 
''delta function'' 
$\delta(C)=(2\pi)^{-1}\int d\lambda\,\exp(i\lambda C)$ 
where $C$ is the wave operator in (\ref{2.2}), 
the integration actually being related to a group average.
Physical states are defined as 
$\psi_{{}_{\rm phys}}=\delta(C)\psi$, where $\psi$ is to
some extent arbitrary 
(in particular, it should not satisfy the wave equation). 
The inner product is given by 
$\langle\psi_{{}_{\rm phys}}|\phi_{{}_{\rm phys}}
\rangle_{{}_{\rm phys}}=\int d\mu\,\psi^* \delta(C)\phi$. 
This naive idea can be made precise in terms of distributions
within the framework provided by the 
''kinematic'' (auxiliary) Hilbert space
$L^2({\cal M},d\mu)$ 
(see also Ref. \cite{Marolf2}). 
The main condition for the procedure to work is that the 
wave operator $C$ is self-adjoint with respect to this
structure. 
\medskip

In contrast to $q$, the Klein-Gordon
type scalar product $Q$ is adequate not only for quantum
cosmology but for the scenario of a
scalar field in a space-time background as well, since it 
does not need information about $ds^2$ and $U$ residing in the 
future of $\Sigma$. It is thus compatible with the idea of
a field $\psi$ propagating causally in $\cal M$. 
On the other hand, such a feature is not at all necessary
in quantum cosmology, and we will transcend it later. 
For the moment, however, our procedure may be interpreted
in terms of quantum cosmology as well as in terms of 
a scalar field in a background. 
\medskip

There is a subtlety related to the question which hypersurfaces
$\Sigma$ may be used in (\ref{2.8}). When both $\psi_1$ and $\psi_2$ 
satisfy the wave equation (\ref{2.2}), the integrand is a
conserved current. Hence, whenever at least one of the two wave
functions vanishes sufficently fast in the asymptotic boundary
region of two spacelike hypersurfaces $\Sigma$ and $\Sigma'$, 
the complex number $Q(\psi_1,\psi_2)$ is independent of whether
it is computed using $\Sigma$ or $\Sigma'$. However, this is
a rather vague statement, since we have not presupposed a particular 
asymptotic structure. One may, for example, think of $\Sigma$
to be a Cauchy hypersurface, and $\Sigma'$ to be a spacelike
non-Cauchy hypersurface (if $ds^2=-d\tau^2+dx^2$, it could
be of the type $\tau^2-x^2=const$). Whether two hypersurfaces
differing so much in their global structure give rise to a unique
scalar product $Q$ will highly depend on the model
--- i.e. on $({\cal M},ds^2,U)$ --- as well as on the space of 
solutions to (\ref{2.2}) admitted. 
The minimal requirement
is that the scalar product $Q$ is compatible with
$\Sigma$ being any of the hypersurfaces $\Gamma_\lambda$
defining the foliation of $\cal M$.
In some models it will thus be necessary to impose additional
asymptotic conditions, e.g. to require all $\Gamma_\lambda$
to be Cauchy hypersurfaces.
We assume that a class of ''admissible hypersurfaces'' has been 
specified so as to fix $Q$ unambigously. Whenever we use a 
spacelike hypersurface
$\Sigma$ to compute $Q$, it shall belong to this class. 
\medskip

In the case of the Klein-Gordon equation in a flat background
($ds^2=-d\tau^2+d\vec{x}^2$, $U=m^2$),
the indefinite nature of $Q$ gives rise to a decomposition
of the space of wave functions $\H$ into positive and negative
frequency subspaces (particle and antiparticle modes), 
$\H=\H^+\oplus\H^-$. The spaces $\H^+$
and $\H^-$ are orthogonal to each other, and $Q$ is positive
(negative) definite on $\H^+$ ($\H^-$). Complex conjugation
maps $\H^\pm$ into each other. In principle, there are
many such decompositions, related to each other by Bogoljubov 
transformations. However, in flat space-time 
the existence of a timelike Killing vector field as well as
the action of the Lorentz group ensures the standard 
(Lorentz invariant)
decomposition to be singled out as {\it preferred} one. 
Moreover, when a Lorentzian coordinate system has been fixed, 
the wave equation reduces to a Schr{\"o}dinger
type equation with unitary evolution within each subspace.
One thus encounters two Hilbert space structures 
$(\H^+,Q)$ and $(\H^-,-Q)$ within each of which an evolution
of the ''ordinary'' quantum mechanics type (although with
non-local Hamiltonian) takes place. Fixing some Lorentzian
coordinate system and writing the wave function as
$\psi(\tau,\vec{x})=\chi(\tau,\vec{x}) e^{-im\tau}$, the
evolution in $\H^+$ (or likewise the definition of
$\H^+$) is given by $i\,\partial_\tau\chi=E \chi$ 
with $E=(m^2-\triangle)^{1/2}-m$. This operator may be well-defined
in the momentum representation (i.e. by means of the 
Fourier transform). The Schr{\"o}dinger type wave
function $\chi$ appears as prefactor of $e^{iS}$ 
with classical action $S=-m \tau$, which in turn is
associated with a congruence of classical particles at rest
in the Lorentzian coordinate system chosen. Since 
this coordinate system was arbitrary, any
congruence of particles in parallel uniform motion provides
a possible ''background'' with respect to which a unitary
evolution equation may be written down. 
It should be remarked that this reasoning corresponds to a
''one-particle interpretation'' and is not the modern way 
to proceed when treating the quantized scalar particle.
However, in quantum cosmology the wave function $\psi$ is usually not
promoted into an operator --- as long as third quantization is not
envisaged --- and one may look at it from the point of view
of a ''one-universe interpretation''. 
We will, in this paper, try to generalize the 
pattern provided by unitarity and decomposition 
to non-trivial backgrounds $({\cal M},ds^2,U)$
as far as possible. 
If one likes in addition the space $\H$ of wave functions to form 
a Hilbert space, one may {\it reverse} the sign of $Q$ in 
$\H^-$ and 
obtain an overall positive definite scalar product. 
Although such a procedure might seem somewhat 
arbitrary, it is completely compatible with the Lorentz symmetry. 
Moreover, the refined algebraic quantization approach
\cite{Aetal,Marolf2},
when applied to the Klein-Gordon equation in Minkowski
space \cite{Marolf1} as well as in a flat space-time with toroidal
spatial sections \cite{Higuchi2} 
(i.e. ${\cal M}=\R\times{\bf S}^1\times{\bf S}^1\times{\bf S}^1$ with
arbitrary radii for the ${\bf S}^1$-factors) precisely yields 
the structure $(\H^+,Q)\oplus (\H^-,-Q)$ 
as the Hilbert space 
$({\cal H}_{{}_{\rm phys}},\langle\,|\,\rangle_{{}_{\rm phys}})$  
of physical states. 
\medskip

A typical way how the conceptual problems in quantum cosmology 
arise is that a generic background $(ds^2,U)$ does not admit
a preferred frequency decomposition of wave functions in an 
obvious way.
This problem persists even when the potential $U$ is positive,
and the notion of outgoing and incoming trajectories (i.e. the
classical analogue of mode decomposition) is unambigously well-defined.
Also, the split of the wave equation into two (mutually
complex conjugate) unitary evolution equations is usually achieved
only in a semiclassical context 
\cite{Vilenkininter} 
where some additional assumption such as the division of
variables into classical and quantum ones 
are necessary (see e.g. Ref. \cite{Kiefersemi} and, 
as a recent contribution touching
upon the problem of unitarity, Ref. \cite{Bertonietal}). 
This type of problem has plagued early attempts to quantize a scalar
particle in curved space-time, 
and similar obstructions have been encountered 
when the question was raised how quantum cosmology should be 
interpreted. 
It might appear that no ''invariant'' structure of the
ordinary quantum mechanical type --- such as Hilbert spaces 
and unitary evolution --- may be constructed at the fundamental 
level 
of the wave equation. This expectation has long been common belief, 
and it relies on the absence of {\it geometric} structures
such as symmetries that would take over the role of the 
timelike Killing vector field in the flat Klein-Gordon 
case \cite{Kuchar1} (see also Ref. \cite{Kuchar2} for the full
superspace context). 
\medskip 

Nevertheless, some efforts have been made to circumvent this 
type of problems, thereby making use of {\it other} than local 
geometric structures. Recently, proposals have been given that
relate a preferred decomposition with the structure of 
superspace at small scale factors \cite{WaldHiguchi} and the 
structure of minisuperspace at large scale factors 
\cite{FE9}. 
Furthermore, the refined algebraic quantization approach 
envisages the construction of a Hilbert space of physical
states quite independent of local geometric 
considerations.
As outlined above, 
it is based on the scalar product $q$ from (\ref{2.7}) 
and thus incorporates global issues
from the outset. It definitely rejects the interpretation of 
wave functions to propagate causally in minisuperspace, and 
hence the use of the Klein-Gordon type product $Q$ from 
(\ref{2.8}). 
Nevertheless, as we have already remarked above, 
its result in the flat Klein-Gordon case consists merely in 
reversing the sign of $Q$ in the 
negative frequency sector. The question whether a similar 
relation exists in more general models has --- to my 
knowledge --- not yet been investigated systematically. 
All these attempts and strategies provide hints that 
{\it in addition} to 
the local geometry of (mini)superspace, some other structure 
may play a fundamental role in the emergence of the 
mathematical objects of quantum mechanics and the possibility of 
predictions. 
\medskip

The aim of this article is to provide further evidence that 
mode decomposition and underlying unitarity might be achieved in a
larger class of models than is usually believed. For the moment,
our main assumption is the positivity of the potential $U$,
i.e. the clear separation of modes in the classical sense. 
After developing the formalism necessary in the next two 
Sections, our proposed solution of the decomposition problem
is formulated in Section 5. There and 
in Section 6 we will encounter hints that the construction 
of an underlying quantum mechanical structure 
along our proposal 
is related to further restrictions,
possibly of the type of {\it analyticity} properties of the model, 
and that it relies on {\it global} issues. In Section 8 we
will speculate on possible relations to the 
refined algebraic quantization approach, and we
give a heuristic argument how to define a preferred
decomposition in this framework 
(although we have to leave open whether this decomposition 
agrees with the one following from our proposal or whether 
it may give rise to a different solution of the decomposition
problem). 
\medskip

\section{Decompositions and unitary evolutions 
associated with WKB-branches}
\setcounter{equation}{0}

Our starting point is to choose a Hamilton-Jacobi function
$S$ (i.e. a solution of (\ref{2.5}) in some
domain $\cal G$) in order to define a geometric background 
structure with respect to which 
the wave equation may further be exploited.
The question whether some structures are actually 
{\it independent} of this choice will be dealt with in 
Sections 4 and 5. 
Hence, for the moment, we consider $S$ to be fixed. Without
loss of generality, we assume $S^\alpha$ to be a future 
(outgoing) oriented vector. Adopting the usual notation, we will
call $S$ an ''action'' function. 
Since we will compute scalar products of the type (\ref{2.8}) with
$\Sigma$ being hypersurfaces of constant $S$, the domain
$\cal G$ shall be of the appropriate type. One may construct
an action function by choosing any spacelike
hypersurface $\Sigma$ admissible for defining the scalar product $Q$
and imposing $S=0$ on $\Sigma$. Then the positivity of $U$
ensures the existence of a solution $S$ at least in a neighbourhood
of $\Sigma$. 
The delicate problem of asymptotic 
spatial boundary conditions for wave functions
will depend on the particular model. 
\medskip

The function $S$ locally generates a congruence of outgoing 
classical trajectories. 
The natural time parameter along these is given by the action 
itself. In order to 
retain the usual convention, we define the time parameter as
\begin{equation}
t=-S\, ,  
\label{3.1}
\end{equation}
future orientation corresponding to increasing values of $t$. 
(This is an immediate consequence of equation (\ref{2.6}), 
taking the pseudo-Riemannian signature of $ds^2$ and the
positivity of the lapse function $N$ into account). 
The evolution parameter $t$ is not indended
to be the physically experienced ''time'', although we take
the freedom to use this word. 
The derivative along the trajctories with respect to
this time parameter shall --- in a sloppy way --- be denoted
as $\partial_t$, and is given by
\begin{equation}
\partial_t =\frac{S^\alpha}{U}\,\nabla_\alpha \, . 
\label{3.2}
\end{equation}
The result of the action of this derivative on a function
$f$ is sometimes written as $\dot{f}$. 
Comparison with equation (\ref{2.6}) shows that this
corresponds to the choice $N=U^{-1}$ of the
lapse function. 
It proves useful to introduce a coordinate system in $\cal G$
adopted to this situation. Let $\xi^a$ ($a=1,\dots (n-1)$)
label the trajectories (i.e. $\dot{\xi^a}=0$) and choose
$(y^\alpha)\equiv (t,\xi^a)$. It then follows that the 
metric is given by  
\begin{equation}
ds^2 = -\,\frac{dt^2}{U} + \gamma_{ab}\,d\xi^a d\xi^b\, , 
\label{3.3}
\end{equation}
where the potential is expressed as a function $U(t,\xi)$, and
$\gamma_{ab}(t,\xi)$ is the Riemannian (i.e. positive definite)
induced metric on the surfaces $\Sigma_t$ of constant $t$ 
(i.e. of constant action). 
In this coordinate system, $\partial_t$ is given by the partial 
derivative  $\partial/\partial t$. 
For any fixed $t$, the functions
$\xi^a$ provide a coordinate system on $\Sigma_t$. For
later purpose let ${\cal D}_a$ denote the covariant derivative
with respect to $\gamma_{ab}$. Again, for fixed $t$, this is an
object intrinsic to the geometry of $\Sigma_t$. 
An $n$-covariant version of $\gamma_{ab}$ is provided by 
\begin{equation}
P_{\alpha\beta}= g_{\alpha\beta} - \, 
\frac{S_\alpha S_\beta}{S^\gamma S_\gamma} \equiv
g_{\alpha\beta} +  
\frac{S_\alpha S_\beta}{U}\, . 
\label{3.4}
\end{equation}
It may also be understood as the projection onto directions
tangential to $\Sigma_t$. 
\medskip

We will encounter linear operators $A$ acting on functions 
$\chi(t,\xi)$ on $\cal M$ whose action is actually tangential 
to $\Sigma_t$ (as e.g. any derivative operator composed
of $\partial_a \equiv\partial/\partial\xi^a$, such as
$A^{ab}(t,\xi)\partial_{ab}$). Such an operator is 
characterized by 
\begin{equation}
[A,S]=0\,  , 
\label{3.5}
\end{equation}
the derivative $\dot{A}$ thereof along the
trajectories being defined as 
the commutator $[\partial_t,A]$. (In the example just mentioned 
this would be $\dot{A}^{ab}(t,\xi)\partial_{ab}$). 
Taking again the commutator with $\partial_t$, it follows that
$[\dot{A},S]=0$, i.e. $\dot{A}$ acts tangential to $\Sigma_t$
along with $A$. 
\medskip

We are now in position to develop our formalism. In addition to
$S$ we fix a real function $D$ satisfying the conservation
equation 
\begin{equation}
\nabla_\alpha (D^2 S^\alpha) = 0\, . 
\label{3.6}
\end{equation}
In terms of the coordinates $(t,\xi^a)$, the general solution
thereof reads
\begin{equation}
D^2(t,\xi)= \frac{f(\xi)}{\sqrt{\gamma(t,\xi)}\,\sqrt{U(t,\xi)}}\,, 
\label{3.7}
\end{equation}
where $f(\xi)$ is an arbitrary function which is constant along
each trajectory and which we assume to be chosen strictly positive 
everywhere in $\cal G$, and $\gamma$ is the determinant of
the metric $\gamma_{ab}$. The function $f$ 
represents the freedom to re-label the trajectories and defines
the measure $d^{n-1}\xi\,f(\xi)$ (see (\ref{3.14}) below). 
\medskip

Given $S$ and $D$, we express any wave function as
\begin{equation}
\psi = \chi D e^{i S} \, . 
\label{3.8}
\end{equation}
This resembles what is usually done in the semiclassical
approach. Note however that we require $S$ and $D$ to satisfy
the full Hamilton-Jacobi and conservation equation, with
all variables $y$ involved. No division of degrees of freedom
into classical and quantum has been assumed. 
Any wave function $\psi$ is associated with a function 
$\chi$ on the domain $\cal G$ in which $S$ and $D$ exist. 
Some wave functions ''fit'' the background structure
$(S,D)$ in a natural way: When
$e^{iS}$ varies much faster than $D$ and $\chi$, $\psi$ may be
called a WKB-state. Although we will not develop a 
WKB type formalism here, we simply call the
pair $(S,D)$ a {\it WKB-branch}. 
(One could equally well call it {\it Hamilton-Jacobi-branch}; 
it encodes a congruence of classical trajectories whose tangent 
vector field is hypersurface orthogonal, endowed with a weight 
represented by $D$. By using the notation {\it admissible} WBK-branch
we emphasize that in a particular model some asymptotic condition 
constraining the set of possible action functions $S$ might be 
appropriate). 
\medskip

Inserting (\ref{3.8}) into (\ref{2.2}), the wave equation
becomes equivalent to
\begin{equation}
i\, \partial_t \chi = \Big(\,\frac{1}{2}\,\partial_{tt} + h \Big)
\chi \, , 
\label{3.9}
\end{equation}
where $h$ is an operator acting tangential to $\Sigma_t$,
i.e. $[h,S]=0$. This may be shown either by direct computation
or, more elegantly, by using an ADM type $(n-1)+1$-decomposition 
of the Laplacian $\nabla_\alpha\nabla^\alpha$\, .
In Ref. 
\cite{FE8} 
the differential equation for $\chi$ has been worked out for 
more general evolution parameters. (Setting
${\cal N}=U^{-1/2}$, ${\cal N}_a=0$ in the formalism of Ref. 
\cite{FE8}
reduces everything to the framework considered here).
We just quote the result that $h$ is given by 
\begin{equation}
h = H^{\rm eff} - \frac{1}{2}  \,D \,(D^{-1})\,\dot{}\,\,\dot{}
\label{3.10}
\end{equation}
where
\begin{equation}
H^{\rm eff} = -\,\frac{1}{2 D U}\,\nabla_\alpha P^{\alpha\beta} 
\nabla_\beta\, D \equiv
-\,\frac{1}{2 D \sqrt{U}}\,
{\cal D}_a \,\frac{1}{\sqrt{U}}\, {\cal D}^a\, D \, .
\label{3.11}
\end{equation}
The derivatives $\nabla_\alpha$ and ${\cal D}_a$ are
understood in this expression to act on everything to the right of
them (e.g. $\nabla_\alpha D$ acting on a function 
$\chi$ giving $\nabla_\alpha(D\chi)$). 
In Ref. \cite{FE8} it was argued that in a WKB-context
(\ref{3.9}) is well-approximated by the effective
Schr{\"o}dinger equation $i\,\partial_t\chi\approx H^{\rm eff}\chi$. 
($H^{\rm eff}$ was therefore called 
''effective Hamiltonian''). 
\medskip

Here we will proceed without approximation instead. 
We first compute the scalar product
(\ref{2.8}) for two wave functions of the form (\ref{3.8}), (i.e. 
within the same WKB-branch $(S,D)$) for 
hypersurfaces $\Sigma=\Sigma_t$. (Due to our 
assumptions, there exists at least an interval of values for $t$ 
such that $\Sigma_t$ is admissible for computing $Q$). 
The orientation of the hypersurface
element is fixed by $d\Sigma^\alpha = d\Sigma \,n^\alpha$, where 
$n^\alpha=-U^{-1/2} S^\alpha$ is the unit normal
($n_\alpha n^\alpha=-1$), and $d\Sigma$ is the (positive definite) 
scalar hypersurface element (in the coordinate system $(t,\xi^a)$ 
we have $n^0<0$ and $d\Sigma = d^{n-1}\xi\,\sqrt{\gamma}\,$). 
Making use of (\ref{3.2}), the definition of $n^\alpha$ and the fact
that $D$ is real, 
the result is easily seen to be 
\begin{equation}
Q(\psi_1,\psi_2) = 
\int_{\Sigma_t} d\Sigma \,\sqrt{U} D^2 
\Bigg(\chi_1^* \chi_2  + \frac{1}{2} 
\Big( \chi_1^* \, (i \partial_t\chi_2) + 
(i \partial_t \chi_1)^* \chi_2 \Big)\Bigg) \, . 
\label{3.12}
\end{equation}
By construction, it is independent of $t$. 
The first part of this expression amounts to introduce the
(positive definite) scalar product for each hypersurface
$\Sigma_t$ within a WKB-branch 
\begin{equation}
\langle\chi_1|\chi_2\rangle\equiv 
\langle\chi_1|\chi_2\rangle_t = 
\int_{\Sigma_t} d\Sigma \, \sqrt{U} D^2 \,\chi_1^*\chi_2 \, .
\label{3.13}
\end{equation}
In the coordinate system $(t,\xi^a)$ this reduces to 
\begin{equation}
\langle\chi_1|\chi_2\rangle = 
\int d^{n-1}\xi\,\sqrt{\gamma} \sqrt{U} D^2 \,\chi_1^*\chi_2 = 
\int d^{n-1}\xi\,f(\xi) \,\chi_1^*(t,\xi)\chi_2(t,\xi) \, .
\label{3.14}
\end{equation}
It thus has the Leibnitz rule property 
\begin{equation}
\frac{d}{dt} \langle\chi_1|\chi_2\rangle = 
\langle\dot{\chi}_1|\chi_2\rangle + 
\langle\chi_1|\dot{\chi}_2\rangle\, . 
\label{3.15}
\end{equation}
In the following we will make use of the formal 
hermitean conjugate $A^\dagger$ of an operator $A$, 
defined by 
\begin{equation}
\langle \chi_1| A \chi_2\rangle = 
\langle A^\dagger \chi_1|\chi_2\rangle \, . 
\label{3.16}
\end{equation}
In case of differential operators, it 
is found by formally carrying out the usual procedures of
partial integration and omitting all terms stemming from the
(asymptotic) boundary of $\Sigma_t$. 
The conditions necessary to make them 
(and statements like $A=A^\dagger$) 
rigorous are addressed to concrete models to which this
formalism is applied. 
The complex conjugate $A^*$ of an operator $A$ is defined
by 
\begin{equation}
A^* \chi = ( A \chi^*)^* \, . 
\label{3.17}
\end{equation}
Complex conjugation satisfies 
$(AB)^*=A^* B^*$, $(A\chi)^*=A^*\chi^*$ and 
$(A^*)^\dagger=(A^\dagger)^*$. 
The operator $h$ from (\ref{3.10}) is hermitean and real 
($h^\dagger=h^*=h$), and the same is true for the effecive
Hamiltonian $H^{\rm eff}$ from (\ref{3.11}) as well as for
the action 
$S$ (as multiplication operator). The latter fact implies that
the hermitan conjugate $A^\dagger$ and the complex conjugate
$A^*$ of an operator $A$ satisfying (\ref{3.5}) both act tangential 
to $\Sigma_t$ along with $A$. 
The dot operation $A\rightarrow \dot{A}\equiv[\partial_t,A]$
commutes with both ${}^*$ and ${}^\dagger\,$. 
\medskip 

The wave equation in the form (\ref{3.9}) is not of the
Schr{\"o}dinger type, but rather a differential equation of
second oder in the classical evolution parameter $t$. 
However, let us try to assume that there exists an operator
$H$ such that any solution of the differential equation 
$i\,\partial_t\chi=H\chi$
automatically satisfies (\ref{3.9}). 
Differentiating this
equation with respect to $t$ and inserting $i\partial_t\chi$ and
$\partial_{tt}\chi$ into (\ref{3.9}), one finds that
the operator 
${\cal C}\equiv i\dot{H} -2h+2H+H^2$ must annihilate all functions
$\chi$ subject to $i\,\partial_t\chi=H\chi$. We would like
in addition that $H$ acts tangential to $\Sigma_t$, which implies
${\cal C}\chi=0$ for all $\chi$ defined on $\Sigma_t$,
hence ${\cal C}=0$. Moreover, a short computation reveals that
the space of all wave functions $\psi$ such that
$\chi$ evolves as $i\,\partial_t\chi=H\chi$ is orthogonal 
under $Q$ to its complex conjugate if 
$H^\dagger=H^*$. The reasoning sketched in these few
lines shall now be reversed. 
\medskip

In view of the brief arguments given above, we write down the
differential equation for an operator $H$, 
\begin{equation}
i \dot{H} = 2 h - 2 H - H^2 \, , 
\label{3.18}
\end{equation}
together with two additional requirements, 
\begin{equation}
[H,S]=0\, . 
\label{3.19}
\end{equation}
and
\begin{equation}
H^\dagger = H^*\, . 
\label{3.20}
\end{equation}
A further requirement will be imposed later (see (\ref{3.33}) below). 
As is easily shown using $[h,S]=0$ and $h^\dagger=h^*$, 
the two constraints (\ref{3.19})--(\ref{3.20}) are propagated 
by the evolution (\ref{3.18}). 
Whenever they are satisfied for one particular value 
$t=t_0$, they are valid for all $t$. Hence,
the system (\ref{3.18})--(\ref{3.20}) is consistent and well-posed. 
\medskip

It is useful to introduce some more notation. 
For any solution $H$ of the system let 
\begin{equation}
H + 1 \equiv {\cal P} \equiv {\cal A}+ i \,{\cal B}\, , 
\label{3.21}
\end{equation}
where ${\cal A}^\dagger={\cal A}^*={\cal A}$, 
${\cal B}^\dagger={\cal B}^*={\cal B}$
and $[{\cal A},S]=[{\cal B},S]=0$. The operators
${\cal A}$ and ${\cal B}$ provide the decomposition 
of  $H+1$ into
its hermitean and anti-hermitean (and at the same time 
into its real and imaginary) part. 
These operators are uniquely determined to be 
\begin{equation}
{\cal A} = 1 + \frac{1}{2} \Big(H + H^\dagger\Big)
\qquad \qquad
{\cal B} =\frac{i}{2}\Big(H^\dagger-H\Big). 
\label{3.22}
\end{equation}
It is in particular the first of these operators that will play 
an important role later on. 
Moreover we note that, if $H$ is a solution, the operator
$\widetilde{H}=-(H^* + 2)$ (which is identical to
$-(H^\dagger+2)$ on account of (\ref{3.20})) is a solution 
as well, the associated quantities being given by 
$\widetilde{{\cal P}}=-{\cal P}^*$, 
$\widetilde{{\cal A}}=-{\cal A}$ and 
$\widetilde{{\cal B}}={\cal B}$. 
\medskip

We assume now that a solution $H$ of 
(\ref{3.18})--(\ref{3.20}) 
has been chosen. Recalling the motivation for considering 
this system of equations, we state the following property: 
Any solution of the equation 
\begin{equation}
i\,\partial_t \chi = H \chi 
\label{3.23}
\end{equation}
gives rise to a solution of the wave equation in the form
(\ref{3.9}), or by virtue of (\ref{3.8}), in the original form
(\ref{2.2}). This follows from the differential
equation (\ref{3.18}). Let us denote the space of wave
functions obtained in this way by $\H^+$. 
Accordingly, 
any solution of the equation 
\begin{equation}
i\,\partial_t \chi = -(H^*+2) \chi 
\label{3.24}
\end{equation}
gives rise to a solution of the wave equation as well.
This is because ${\widetilde H}=-(H^*+2)$ satisfies the
differential equation (\ref{3.18}) along with $H$.  
The space of wave functions obtained in this way shall be
denoted by $\H^-$. 
We have thus defined two subspaces $\H^\pm$ of the total space
$\H$ of wave functions (whereby 
we assume appropriate fall-off conditions to hold at the 
boundaries of $\Sigma_t$, if necessary). 
\medskip

Turning to the question how these three spaces are related
to each other, we observe that $\H^\pm$ are complex
conjugate to each other, i.e. $(\H^\pm)^* = \H^\mp$. 
In order to see this it is important to recall that we have
chosen a WKB-branch $(S,D)$ with respect to which {\it any}
wave function $\psi$ is traced back to a function
$\chi$. The relation between $\psi$ and $\chi$ is always
provided by (\ref{3.8}). Hence, if the complex conjugate
$\psi^* = \chi^* D e^{-iS}$ 
of a wave function $\psi$ shall be represented in terms of
the WKB-branch $(S,D)$, it must be rewritten as
$\psi^* = \chi_c D e^{i S}$, with 
$\chi_c=\chi^* e^{-2iS}\equiv \chi^* e^{2it}$ being
the associated function. 
(In contrast, the original expression
$\psi^* = \chi^* D e^{-iS}$ refers to the
WKB-branch $(-S,D)$ instead). 
Complex conjugation $\psi\rightarrow\psi^*$ of wave functions 
is thus represented as $\chi\rightarrow\chi_c$. It is now
an easy task to show that $\chi$ satisfies the evolution
equation (\ref{3.23}) if and only if $\chi_c$ satisfies
(\ref{3.24}). The number $2$ in the operator
$-(H^*+2)$ thus represents the separation between the 
two WKB-branches $(\pm S,D)$ by a total relative phase factor
$e^{-2 i S}$. 
This can be illustrated by a heuristic argument: 
In a semiclassical (WKB) context one may expect $H$ to be
''small'' as compared to unity, and thus 
(\ref{3.23}) to describe a slowly varying function $\chi$.
As a consequence, (\ref{3.8}) is dominated by the phase
factor and represents a WKB type state. Its complex 
conjugate is a WKB type state as well (with respect to
the WKB-branch $(-S,D)$, i.e. with reversed trajectories), but 
the associated function $\chi_c$ --- the representation of
$\psi^*$ in the WKB-branch $(S,D)$ --- is now rapidly
varying!
\medskip

Since the two spaces $\H^\pm$ are complex conjugate to each
other, whatever
can be stated about $\H^+$ and the evolution equation
(\ref{3.23}) has a counterpart in terms of $\H^-$ and 
the evolution equation (\ref{3.24}). 
It is thus natural to ask whether $\H^+$ and $\H^-$ have a
non-trivial intersection, and whether their sum spans
the space $\H$ of all wave functions. The answer to these
questions depends on the properties of the operator $\cal A$. 
It proves useful to insert (\ref{3.21}) into (\ref{3.18}).
The differential equation
for $H$ becomes equivalent to 
\begin{equation}
i\, \dot{\cal P} + {\cal P}^2 =1+2h\, . 
\label{3.25}
\end{equation}
Separating real and imaginary parts, we find the system  
\begin{eqnarray}
\dot{\cal A} &=& -\, \{{\cal A},{\cal B}\}
\label{3.26}\\
\dot{\cal B} &=& {\cal A}^2 -{\cal B}^2 -1-2h\, , 
\label{3.27}
\end{eqnarray}
where $\{{\cal A},{\cal B}\}$ denotes the anticommutator
${\cal A}{\cal B} + {\cal B}{\cal A}$. 
(Let us note {\it en passant} that yet another equivalent
form of (\ref{3.18}) is achieved by setting 
${\cal P}= i\,\dot{u}\,u^{-1}\,$. The resulting equation for 
the operator $u$ is linear,
\begin{equation}
\ddot{u}+(1+2h)\,u=0\,,
\label{3.28} 
\end{equation}
which is not a surprise since $H$ stems from a linear problem. We 
will make use of this result later on). 
Equation (\ref{3.26}) may formally be solved: Let ${\cal A}_0$ 
be the action
of $\cal A$ on some ''initial'' hypersurface $\Sigma_{t_0}$. 
Since $\cal A$ acts
tangential to all $\Sigma_t$ (i.e. $[{\cal A},S]=0$),
one may imagine to represent $\cal A$ as a $t$-dependent
operator acting on functions $\chi(\xi)$, and then 
insert $t=t_0$. This makes ${\cal A}_0$ an operator for functions
$\chi(y)$, i.e. at the same footing as all other operators
considered so far. The solution to (\ref{3.26}) is given by
\begin{equation}
{\cal A} = {\cal U} {\cal A}_0\, {\cal U}^\dagger
\label{3.29}
\end{equation}
where $\cal U$ is the operator satisfying the differential 
equation 
$\dot{\cal U}=-{\cal B}\,{\cal U}$ and the initial condition
${\cal U}_0 = 1$. Its hermitean conjugate thus satisfies
$\dot{\cal U}^\dagger=-{\cal U}^\dagger {\cal B}$, and
(\ref{3.29}) is easily seen to be the unique solution to
(\ref{3.26}), once $\cal B$ is known. 
Hence, whenever there is a function $\chi_0$ on some $\Sigma_{t_0}$
such that ${\cal A}_0 \chi_0=0$ on $\Sigma_{t_0}$, it follows
that there is a function $\chi$ in the region on which
$\cal A$ exists, such that ${\cal A}\chi=0$. 
This ensures that there exist solutions $H$ such that
${\cal A}$ is invertible (or positive). As an example,
one may choose $H_0=0$. 
\medskip

There is another way to make this explicit. If $\psi\in \H^+$,
i.e. $\chi$ satisfies (\ref{3.23}), then
$\mu={\cal A}\chi$ satisfies $i\partial_t\mu=H^\dagger\mu$.
Likewise, if $\psi\in \H^-$,
i.e. $\chi$ satisfies (\ref{3.24}), then
$\mu={\cal A}\chi$ satisfies $i\partial_t\mu=-(H+2)\mu$.
If ${\cal A}_0\chi_0=0$ on a hypersurface $\Sigma_{t_0}$,
we promote $\chi_0$ into a solution of (\ref{3.23}). Then
$\mu={\cal A}\chi$ vanishes on $\Sigma_{t_0}$ and
satisfies a Schr{\"o}dinger type equation. As a
consequence $\mu=0$, i.e. ${\cal A}\chi=0$ everywhere.
In other words, the existence of functions annihilated by
$\cal A$ cannot be restricted to a single
hypersurface $\Sigma_{t_0}$. In this sense, the
character of $\cal A$ as an invertible (or positive) operator
is propagated by the evolution provided by (\ref{3.18}). 
\medskip

Subtracting (\ref{3.24}) from (\ref{3.23}), we infer that
a common element of both $\H^+$ and $\H^-$ satisfies
${\cal A}\chi=0$. We require that this equation has no 
non-trivial solution. 
In a particular model, 
one must refer to the asymptotic boundary conditions defining $\H$ 
in order to exclude the existence of a non-trivial $\chi$. 
We assume that this can be done and end up with
$\H^+\cap\H^- =\{0\}$. 
Consequently, one may consider the direct sum
$\H^+\oplus\H^-$ and ask whether it coincides with $\H$. 
Given any $\psi\in \H$, we consider its ''initial data''
$(\chi,\dot{\chi})$ at some hypersurface $\Sigma_{t_0}$. 
$\psi$ is the sum of two wave functions $\psi^\pm\in\H^\pm$ 
if the system 
of equations $\chi^+ + \chi^- = \chi$,
$H\chi^+ -(H^* +2)\chi^- =i\,\dot{\chi}$
may be solved with respect to $\chi^\pm$ on some $\Sigma_{t_0}$.
The propagation of $\chi^\pm$ off $\Sigma_{t_0}$ 
is then provided by (\ref{3.23}) and (\ref{3.24}).
Eliminating $\chi^+$ from these equations, one arrives at
the single equation 
$2 {\cal A}\chi^- = (H-i\,\partial_t)\chi$, which has to 
be solved with respect to $\chi^-$ on $\Sigma_{t_0}$. 
Since $\chi$ and
$\dot{\chi}$ may arbitrarily be prescribed on $\Sigma_{t_0}$
(up to the required fall-off behaviour at the asymptotic boundary),
the decomposition of $\psi$ into a sum of $\psi^\pm$ 
may always be achieved if ${\cal A}$ has
a sufficiently well-behaved inverse. 
\medskip

The structure of the scalar product for elements of $\H^\pm$ 
depends on the properties of $\cal A$ as well. 
If $\psi_{1,2}\in\H^+$, a straightforward use of 
(\ref{3.12}) reveals
\begin{equation}
Q(\psi_1,\psi_2) = \langle\chi_1|{\cal A}\chi_2\rangle\, , 
\label{3.30}
\end{equation}
whereas for $\psi_{1,2}\in\H^-$ we find
\begin{equation}
Q(\psi_1,\psi_2) = -\langle\chi_1|{\cal A}\chi_2\rangle\, . 
\label{3.31}
\end{equation}
If $\psi_1\in \H^+$ and $\psi_2\in\H^-$, the scalar
product is 
\begin{equation}
Q(\psi_1,\psi_2) = \frac{1}{2}\, 
\langle\chi_1|(H^\dagger-H^*)\chi_2\rangle \, , 
\label{3.32}
\end{equation}
which vanishes on account of (\ref{3.20}).
\medskip

In view of these results we impose an 
additional requirement on the solutions $H$ of
the system (\ref{3.18})--(\ref{3.20}). We assume that
$\cal A$ is a positive operator, symbolically
\begin{equation}
{\cal A}>0\, .
\label{3.33}
\end{equation}
As a consequence, 
the Klein-Gordon type scalar product $Q$ is positive 
(negative) definite on $\H^+$ ($\H^-$), and the total space
of wave functions is a direct orthogonal sum
\begin{equation}
\H=\H^+\oplus\H^- \, .
\label{3.34}
\end{equation}
Thus, if all steps we have presented
in a rather formal way go through in a particular model, any
choice of a WKB-branch $(S,D)$ and any
choice of $H$ subject to the above restrictions
gives rise to two Hilbert spaces of wave functions. 
Note that for this construction to fix $\H^\pm$ uniquely,
it is sufficient that the basic objects, $S$ and $H$, exist in a 
neighbourhood of some hypersurface $\Sigma_{t_0}$. 
The key role of $H$ is thus to fix the initial data of $\psi^\pm$. 
At the boundary of the domain on which $S$ and $D$ exist, 
the relation (\ref{3.8}) between $\psi$ and $\chi$ 
breaks down, thus leading to a singular behaviour of $\chi$ 
and $H$, while $\psi$ remains well-behaved. 
\medskip

We have already mentioned that in a WKB-context the operator $H$
can be considered as much ''smaller'' than unity. In such a context,
a solution of (\ref{3.23}) may be viewed to be a WKB type state
''built'' mainly 
around outgoing trajectories. Analogously, a solution of 
(\ref{3.24}) may be viewed to be associated with incoming
trajectories. Thus, the spaces $\H^\pm$ provide generalizations of
the classical modes of ''outgoing'' and ''incoming'' evolution to 
the quantum level. In this sense, we can talk about a ''mode''
decomposition. Note however that, so far, there is considerable freedom
in defining such a decomposition.  
In Section 5 we will examine the perspectives for making it unique.
\medskip 

The practical use we will now make of the positivity property of 
$\cal A$ is that there exists a positive square root 
${\cal A}^{1/2}$ and its inverse ${\cal A}^{-1/2}$. 
So far, 
we have constructed two Schr{\"o}dinger type equations 
(\ref{3.23})--(\ref{3.24}), but the evolution operators 
$H$ and $-(H^*+2)$ 
are in general not hermitean. This may easily be cured by 
redefining the 
wave functions once more. If the WKB-branch $(S,D)$ is fixed,
equation (\ref{3.8}) defines a one-to-one correspondence
between $\psi$ and $\chi$. Having chosen $H$, we define
for any wave function 
\begin{equation}
\eta = {\cal A}^{1/2}\,\chi \, . 
\label{3.35}
\end{equation}
For $\psi_{1,2}\in\H^+$ we thus find
\begin{equation}
Q(\psi_1,\psi_2) = \langle\eta_1|\eta_2\rangle\, , 
\label{3.36}
\end{equation}
whereas for $\psi_{1,2}\in\H^-$ we have 
\begin{equation}
Q(\psi_1,\psi_2) = - \langle\eta_1|\eta_2\rangle\, . 
\label{3.37}
\end{equation}
For completeness we recall that $Q(\psi_1,\psi_2)=0$ if
$\psi_1\in\H^+$ and $\psi_2\in\H^-$. 
Since (\ref{3.35}) applies for any state, 
the decomposition of a given wave function with respect
to $\H^\pm$ reads $\eta=\eta^+ + \eta^-$. 
\medskip

The evolution equations (\ref{3.23})--(\ref{3.24}) may now easily 
be re-written in terms of $\eta$. 
If $\psi\in\H^+$ we find 
\begin{equation}
i\,\partial_t \eta ={\cal K}\eta\,, 
\label{3.38}
\end{equation}
which replaces (\ref{3.23}), and for $\psi\in\H^-$ we have 
\begin{equation}
i\,\partial_t \eta =-({\cal K}^* +2)\eta \,, 
\label{3.39}
\end{equation}
which replaces (\ref{3.24}). 
The operator ${\cal K}$ appearing here is given by 
\begin{equation}
{\cal K} = {\cal A}^{1/2} H {\cal A}^{-1/2} + 
i \,({\cal A}^{1/2})\,\dot{} \, {\cal A}^{-1/2}\, . 
\label{3.40}
\end{equation}
It is hermitean, 
\begin{equation}
{\cal K}^\dagger ={\cal K}\,, 
\label{3.41}
\end{equation}
and acts tangential to $\Sigma_t$, i.e. 
$[{\cal K},S]=0$. Note that 
in general ${\cal K}^\dagger\neq {\cal K}^*$. 
The same properties hold for the evolution operator of (\ref{3.39}), 
in particular $-({\cal K}^* +2)^\dagger =-({\cal K}^* +2)$. 
(The way how properties of ${\cal K}$ carry over to
properties of $-({\cal K}^* +2)$
may be inferred analogously to the case of the operators
$H$ and $-(H^* +2)\,$:
If a wave function $\psi$ is represented in the WKB-branch $(S,D)$
by the function $\eta$,
then the complex conjugate wave function $\psi^*$ is represented
by $\eta_c =\eta^* e^{-2iS}$). 
By using (\ref{3.21}), (\ref{3.26}), 
the time derivative of the identity 
${\cal A} = {\cal A}^{1/2} {\cal A}^{1/2}$, and the
fact that $({\cal A}^{1/2})^\dagger={\cal A}^{1/2}$, 
we can write down another expression for $\cal K$, 
\begin{equation}
{\cal K}= {\cal A}-1 + \frac{i}{2}\,
{\cal A}^{-1/2}
\Big( 
[{\cal A},{\cal B}] + 
[{\cal A}^{1/2},({\cal A}^{1/2})\,\dot{}\,\,]\Big) 
{\cal A}^{-1/2} \, , 
\label{3.42}
\end{equation}
which makes its hermiticity property (\ref{3.41}) explicit. 
In this formula, one may insert 
$[{\cal A},{\cal B}] = \frac{i}{2}[H,H^\dagger]$. 
\medskip 

Both (\ref{3.38}) and (\ref{3.39}) 
represent (at least at the formal way we are
treating things here) a unitary evolution 
with respect
to the (negative of the) classical action $S$ 
as time parameter. The coordinates
$\xi^a$ labeling the trajectories play the role
of standard quantum mechanical variables. 
For a given wave function $\psi(y)$, 
the function $\eta(t,\xi)$ may thus in a sense be considered as the
''true'' representation of the quantum state with respect 
to the background structure $(S,D,H)$ .
\medskip 

\section{Variations between WKB-branches}
\setcounter{equation}{0}

The WKB-branch $(S,D)$ as well as the operator $H$ used in the previous 
Section have not yet been specified but have been chosen
arbitrarily. 
Here, we will examine what happens when a wave function is
analyzed with respect to two WKB-branches. 
\medskip

Let $(S,D)$ and $(S',D')$ be two infinitesimally close WKB-branches. 
They can be assumed to be defined in the same domain, which thus
contains a family of hypersurfaces $\Sigma_t$ a well as a family of
hypersurfaces $\Sigma_{t'}\,$. One may imagine these two families
of hypersurfaces to be small deformations of each other. 
We will denote variations at fixed points of the manifold
$\cal M$ by $\delta S = S'-S$, $\delta D = D'-D$ etc. 
The background quantities are of course unchanged (in
particular $\delta U=0$, and $\delta$ commutes with
$\nabla_\alpha$). 
Finite variations can in principle be constructed 
from infinitesimal ones, as long as the initial and the
final WKB-branch are not separated by some barrier of 
global type. 
\medskip

The variation is characterized by $\delta S$ and $\delta D$,
which have to respect the Hamilton-Jacobi (\ref{2.5}) as well as 
the conservation equation (\ref{3.6}). These quantities are easily
seen to be constrained by
\begin{eqnarray}
(\delta S)\,\dot{} &=& 0
\label{4.1}\\
\Big(\frac{\delta D}{D}\Big)
\begin{array}{ccc} {\!\!\!\!\dot{}}\\ {} 
\end{array}\!\!\!\!\! &=&
-\,\frac{1}{2\,U D^2}\, 
\nabla_\alpha (D^2\delta S^\alpha)\, . 
\label{4.2}
\end{eqnarray}
It is convenient to write down these objects in terms
of the original (''unperturbed'') WKB-branch $(S,D)$. 
In a coordinate system $(y^\alpha)\equiv (t,\xi^a)$, the 
first equation 
states that $\delta S(\xi)$ is an arbitrary function, and
$\delta D(t,\xi)$ can be found by integration along the trajectories
of $S$, the freedom to choose the initial value $\delta D(t_0,\xi)$ 
corresponding to the expected 
freedom for $D'$. 
The variation of the derivative operator (\ref{3.2}) is given
by
\begin{equation}
\delta(\partial_t) \equiv \mbox{\boldmath $\nabla$}
= \frac{\delta S^\alpha}{U}\,\nabla_\alpha \, . 
\label{4.3}
\end{equation}
This is a derivative operator purely tangential to $\Sigma_t$ 
(i.e. $[\mbox{\boldmath $\nabla$},S]=0$)  
illustrating the change in the direction of the trajectories
(and thus the variation of the hypersurfaces of constant action). 
\medskip

Since the operator $h$ from (\ref{3.10}) is uniquely defined in
each WKB-branch, its variation is already fixed. 
Neverthless, straightforward application of $\delta$ gives
an awkward expression that may be simplified in a somewhat
lengthy computation. On that way, one uses identities like
$P^{\alpha\beta}\delta S_\beta=\delta S^\alpha$, 
$S_\beta \delta P^{\alpha\beta} = - \delta S^\alpha$ and
$[\delta h,S]=-[h,\delta S]$ (which follows from the fact
that $[h,S]=0$). After performing some operator algebra
involving (anti)commutators with $\partial_t$, one may get
rid of several inconvenient terms stemming from the second
contribution in (\ref{3.10}) and containing derivatives of $D$. 
Denoting
\begin{equation}
{\cal E} = \frac{\delta D}{D}+ i\,\delta S \, , 
\label{4.4}
\end{equation}
the result is
\begin{equation}
\delta h = [h,{\cal E}] + 
i \, ( \mbox{\boldmath $\nabla$} - \dot{{\cal E}}) 
-\,\frac{1}{2}\,[\partial_t, \mbox{\boldmath $\nabla$} - \dot{{\cal E}}] 
+[h,\delta S]\,\partial_t\, . 
\label{4.5}
\end{equation}
Due to (\ref{4.1}), one may insert 
$\dot{{\cal E}}=(D^{-1}\delta D)\,\dot{}\,\,$. 
Another expression for this variation is 
\begin{equation}
\delta h = [h,\frac{\delta D}{D}] + \,\frac{1}{2}\, 
[\dot{h},\delta S] + [h,\delta S]\,\partial_t \, ,
\label{4.6}
\end{equation}
where one may replace $[\dot{h},\delta S]$ by
$[\partial_t,[h,\delta S]\,]\equiv [h,\delta S]\,\dot{}\,$. 
Also note that 
$[h,\delta S]=-\mbox{\boldmath $\nabla$} +\dot{{\cal E}}$. 
This form is particularly suitable for computing the variation
of $\dot{h}$ and higher derivatives. 
The most beautiful and comprimed formula is 
\begin{equation}
\delta h = [h,\frac{\delta D}{D}] + \,\frac{1}{2}\, 
\{\partial_t, [h,\delta S]\}\, . 
\label{4.7}
\end{equation}
In all these expressions, the appearance of $\partial_t$ as 
an operator acting directly on wave functions 
signals the variation of the trajectories. Although
$h'$ contains only derivatives with respect to the coordinates
$\xi'{}^a$, it attains derivatives $\partial_t$ when
re-written in terms of $(t,\xi^a)$. 
\medskip

Having made explicit how the basic quantities defining a
WKB-branch transform, we turn to the case of $H$. In both
WKB-branches, we have the system of equations 
(\ref{3.18})--(\ref{3.20}) for $H$ and $H'$, respectively. 
In principle, both $H$ and $H'$ may be chosen completely
independent of each other, as long as they satisfy their
respective equations (we just assume that they differ only by
an infinitesimal operator $\delta H$). One may of course take the
variation of (\ref{3.18}) --- thereby using
$\delta\dot{H}\equiv\delta([\partial_t,H]) =
(\delta H)\,\dot{} + [\mbox{\boldmath $\nabla$},H]\,\,$ --- 
in order to obtain a linear inhomogeneous differential
equation for $\delta H$ that has to be satisfied
anyway. (It states that, if $H$ obeys (\ref{3.18}),
then $H'$ obeys its respective differential equation as well). 
Also, one may vary (\ref{3.19}) and (\ref{3.20}), whereby
in the latter case one must take into account 
the subtlety that 
$\delta$ does not commute with ${}^\dagger\,$. 
\medskip

We will however restrict the freedom in choosing $H$ and
$H'$ independently. We require that {\it the mode decompositions}
induced by these two operators {\it are identical}, i.e.
$\H^+ = \H'^+$ and $\H^- = \H'^-\,$. Any wave function can
be re-written in terms of the two WKB-branches as
$\psi=\chi D e^{iS}=\chi'D'e^{iS'}$. If $\psi$ is an element
of both $\H^+$ and $\H'^+$, we encounter two evolution
equations $i\,\partial_t\chi=H\chi$ and
$i\,\partial_{t'}\chi'=H'\chi'$. 
Expressing $\chi$ and $\partial_t$ in terms
of $\chi'$ and $\partial_{t'}$ and assuming 
$\delta H \equiv H'- H$ to be infinitesimal,
a simple calculation reveals that the variation of $H$ must 
be of the form 
\begin{equation}
\delta H = [H,{\cal E}] + 
i \, ( \mbox{\boldmath $\nabla$} - \dot{{\cal E}}) 
+ {\cal O}\, , 
\label{4.8}
\end{equation}
where $\cal E$ is given by (\ref{4.4}) and 
$\cal O$ is an operator satisfying ${\cal O}\chi=0$ 
for all $\chi$ such that $i\,\partial_t \chi=H\chi$. 
Taking into account that (\ref{3.19}) is valid in both 
WKB-branches (hence $[\delta H,S]+[H,\delta S]=0$), 
it follows that 
${\cal O}=[H,\delta S](\partial_t + i H)$. 
Thus, $\delta H$ is completely fixed, once $H$ has been chosen. 
(Supposing that ${\cal O}$ has an additional contribution
${\cal O}^{\rm add}$, we must have 
$[{\cal O}^{\rm add},S]=0$ and 
${\cal O}^{\rm add}\chi=0$ for all $\chi$ such that
$i\,\partial_t \chi=H\chi$. Since 
${\cal O}^{\rm add}$ acts tangential to $\Sigma_t$, and
the ''initial value'' of $\chi$ on some $\Sigma_{t_0}$ 
is arbitrary, we infer that ${\cal O}^{\rm add}\chi=0$ 
for all $\chi$, hence ${\cal O}^{\rm add}=0$). 
In other words, $H$ and $H'$ must be such that their
(infinitesimal) difference is given by 
\begin{equation}
\delta H = [H,{\cal E}] + 
i \, ( \mbox{\boldmath $\nabla$} - \dot{{\cal E}}) 
+ [H,\delta S]\,(\partial_t + i H)\, . 
\label{4.9}
\end{equation}
Inserting $\mbox{\boldmath $\nabla$} -\dot{{\cal E}} =- [h,\delta S]$,
we obtain 
\begin{equation}
\delta H = [H,\frac{\delta D}{D}] 
+i\,[H-h,\delta S] + [H,\delta S]\,(\partial_t + i H)\, . 
\label{4.10}
\end{equation}
This condition is equivalent to the equality $\H^+ = \H'^+$. 
Complex conjugation implies the equality 
$\H^- = \H'^-$. 
\medskip

We have now a simple criterion whether the decompositions defined
in terms of two overlapping WKB-branches are equal. 
Our idea is that the whole of the manifold $\cal M$ is
covered by all admissible WKB-branches of the previous type.
Furthermore we assume that this set of admissible WKB-branches
is connected in the sense that any two branches may be deformed
into each other by a family of infinitesimal variations.
Whether this idea may rigorously be formulated will depend on the
details of the particular model, its asymptotic structure
and on how ''admissible'' is defined. Here, we assume that 
it is the case. 
\medskip

As a consequence, one could try to choose 
the operator $H$ within any WKB-branch  
such that for infinitesimal neighbours
the condition (\ref{4.10}) holds. This in turn gives rise to
a unique decomposition. One may, for example, start with 
any decomposition of $\H$ and define $H$ in each WKB-branch 
such that it generates the pre-assumed decomposition. Then 
(\ref{4.10}) will surely hold for close neighbours. This
argument shows the existence of such constructions.
However, in doing so, nothing would be gained. 
Our actual goal is to single out a {\it preferred} decomposition.
\medskip

At this step we must make a difference between the two 
possible interpretations
of the wave equation (\ref{2.1}) as a quantized scalar particle and
as a quantum cosmological model. In the first interpretation, the idea 
of a wave function propagating causally is natural if not necessary. 
This provides the main motivation for using the 
Klein-Gordon type scalar product $Q$ rather than the 
$L^2$-product $q$ from (\ref{2.7}) in quantum field theory
on curved space \cite{BirrellDavies}. 
On the other hand, no such feature is necessary in quantum cosmology. 
When trying to select a solution of the differential equation
(\ref{3.18}), we will accept a result that lacks 
causality and locality properties and thus transcends
the quantized particle picture. 
\medskip

\section{Perspectives for uniqueness: the iterative solution}
\setcounter{equation}{0}

In this Section we will tackle the problem whether a 
preferred decomposition $\H=\H^+\oplus\H^-$ may be defined 
in a natural way. 
According to the formalism developed in the two foregoing 
Sections, we have to choose an operator $H$ within each admissible
WKB-branch $(S,D)$, satisfying the differential equation
(\ref{3.18}) and the additional 
conditions (\ref{3.19})--(\ref{3.20}). For any two 
infinitesimally close WKB-branches, the variation shall 
satisfy the equation (\ref{4.10}). This guarantees that the
decompositions associated with the two WKB-branches are identical. 
As already mentioned, we assume the global structure of the set
of all admissible WKB-branches to imply the statement that, 
no matter which WKB-branch is used to define the decomposition, 
the result is alway the same. 
\medskip

Thus, the next step in our program is to make an
appropriate choice of the operator $H$, once $(S,D)$ are given. 
This choice has to be ''natural'', and should thus not 
refer to additional objects (such as a prescribed
decomposition). Since (\ref{3.18}) is a differential equation, 
one might think of fixing the ''inital value'' of $H$ on some
hypersurface $\Sigma_{t_0}$. Two possible guesses might be $H=0$ or
$H=h$ at $t=t_0$. However, a simple computation reveals that
the solutions of (\ref{3.18}) defined by these choices cannot satisfy
(\ref{4.10}). The reason is that two WKB-branches differing
merely by a rigid translation $t\rightarrow t+const$ must be
considered different (with $\delta S=const$). 
In other words, the value $t=t_0$ does not have an invariant 
significance, and any choice of a solution $H$ based on
initial conditions will fail. Let us add that the attempt 
to use initial conditions for selecting $H$ is in spirit related 
to the scalar particle quantization picture of our problem. 
The impossibility in doing so reflects the fact that 
no preferred decomposition may be singled out in a 
causal and local way. This may be viewed as a justification 
of the point of view that a unique mode decomposition is
not the appropriate thing to look for in quantum field theory
in a curved background. However, we are mainly interested in 
quantum cosmology, where new possibilities arise. 
\medskip

We will now present a formal solution to the problem. It is 
''formal'' in the sense that a power series emerges whose actual
convergence (possibly in some regularized sense) may depend on the 
particular model. The procedure described in the following is thus
understood as a general strategy whose application to concrete
models or classes of models will raise a number of questions by
its own. 
\medskip

The starting point is to rewrite 
(\ref{3.18}) as
\begin{equation}
H = h -\,\frac{1}{2}\, H^2-\,\frac{i}{2}\,\dot{H} \, . 
\label{5.1}
\end{equation}
This equation may be iterated, i.e. the two terms
$H^2$ and $\dot{H}$ may be replaced by the square and the
time-derivative of 
$h -\,\frac{1}{2}\, H^2-\,\frac{i}{2}\,\dot{H}$, followed
by the same substitution whenever $H$ appears in this
new equation, and so forth. One obtains an expression
\begin{equation}
H = h -\,\frac{1}{2}\, h^2-\,\frac{i}{2}\,\dot{h} + 
\frac{1}{2}\, h^3 +\frac{i}{2}\,\{h,\dot{h}\}-\, 
\frac{1}{4}\, \ddot{h} + \dots
\label{5.2}
\end{equation}
in which the dependence of the rhs on $H$ has disappeared. 
Only expressions involving $h$ and its derivatives remain. 
Whenever this series converges (in a suitable topology on the 
space of operators) it defines a
solution of (\ref{5.1}). Also, the constraints 
(\ref{3.19})--(\ref{3.20}) will be satisfied on account of 
the properties of $h$ and the operations ${}^\dagger$, 
${}^*$ and $\partial_t\,$. Surprisingly, this solution --- if it
exists --- solves the decomposition problem, as we will now show. 
\medskip

When iterating (\ref{5.1}), it is useful to introduce a formal
book-keeping parameter $\epsilon$ whose actual value is $1$, 
and which keeps track of the ''order'' of terms appearing in
(\ref{5.2}). By replacing 
$h\rightarrow\epsilon h$, $H\rightarrow\epsilon H$ and 
$\partial_t\rightarrow\epsilon \partial_t$, the differential
equation (\ref{5.1}) becomes 
\begin{equation}
H = h -\,\frac{1}{2}\,\epsilon\, H^2- 
               \,\frac{i}{2}\,\epsilon\,\dot{H} \, . 
\label{5.3}
\end{equation}
This reflects the structure of the iteration and allows to think of
$\epsilon$ as a ''small'' expansion parameter. 
The rescaling of quantities by insertion of appropriate powers of
$\epsilon$ could have been achieved from the outset by using
the evolution parameter $t^{\rm new}=\epsilon t\equiv-\epsilon S$ 
and the operators 
$h^{\rm new}=\epsilon^{-1} h$, 
$H^{\rm new}=\epsilon^{-1} H$ instead of $t$, $h$ and $H$, and
thereafter removing the superscript ${}^{\rm new}\,$. 
As a consequence, the according insertions in any of the
formulae displayed in the following is unique and 
consistent. This consistency is crucial for our construction. 
Note that we do {\it not} introduce an $\epsilon$-dependence 
of $h$, as the second part of equation (\ref{3.10}) 
might suggest. Thus, our particular way of inserting 
$\epsilon$'s {\it differs} from a mere book-keeping of an 
$\hbar$-type parameter, and it is actually simpler. 
It is designed such that in the procedure given below
the operator $h$ may be treated as an object by its own, 
without any reference to its internal structure, i.e. to 
its definition (\ref{3.10})--(\ref{3.11}). 
Nevertheless, the position of $\epsilon$ in (\ref{5.3}) indicates 
Planck scale quantities (such as $m/m_P$, where $m$ is a
mass scale in the matter sector of the theory) 
in a realistic model. In a semiclassical context one would 
expect $H\approx h$ (in fact $H\approx h\approx H^{\rm eff}$ 
because the second term in (\ref{3.10}) is smaller than
the first one by a Planck-scale order 
\cite{FE8}, a fact that is ignored by our way to insert
$\epsilon$). In this sense, $\epsilon$ may in fact be 
expected to trace ''small'' contributions, 
although quantities like $h$ contain ''small'' parts as well
which are not indicated by $\epsilon$. The simplicity of
our book-keeping scheme precisely relies on this subtlety. 
\medskip

We remark here for completeness that the insertion of $\epsilon$'s 
along the lines described above changes the the wave equation 
(\ref{3.9}) into 
$i\partial_t \chi=(\frac{1}{2}\epsilon\partial_{tt}+h)\chi$, 
whereas the definition (\ref{3.21}) becomes
$\epsilon H+1\equiv{\cal P}\equiv{\cal A}+i{\cal B}$. 
The formula (\ref{3.40}) for the operator ${\cal K}$ 
as well as the unitary evolution equations
(\ref{3.38})--(\ref{3.39}) remain as they stand (if one rescales 
${\cal K}\rightarrow\epsilon{\cal K}$). 
Equation (\ref{3.25}) takes the form 
$i \epsilon\dot{\cal P} + {\cal P}^2 = 1+2\epsilon h$. 
The linear differential equation (\ref{3.28}) will be
re-written below (equation (\ref{5.16})). 
\medskip

Returning to (\ref{5.3}), 
the iteration procedure generates a power series
\begin{equation}
H = \sum_{p=0}^\infty {\cal H}_p \epsilon^p \, , 
\label{5.4}
\end{equation}
by which we mean that, for any given order $p$, a finite number 
(actually $p+1$) of
iterations generates the operators 
${\cal H}_0,\dots {\cal H}_p\,$. Any further iteration leaves
these operators unchanged (i.e. acts only on
${\cal H}_{p+1},{\cal H}_{p+2}\dots\,$). When the
iteration procedure leads to a convergent result, it should be
identical to (\ref{5.4}), evaluated at $\epsilon=1$. 
There are two ways to formulate this situation in terms of
sequences of well-defined operators. The first one is to consider
the iterative scheme
\begin{eqnarray}
H_0 &=& 0\nonumber\\
H_{p+1} &=& h -\,\frac{1}{2}\, \epsilon \,H_p^2 -\, 
        \frac{i}{2}\,\epsilon\, \dot{H}_p
\label{5.5}
\end{eqnarray}
for non-negative integer $p$. After an appropriate number of
steps, the operators ${\cal H}_p$ may be read off. The iterative
solution to the differential equation (\ref{5.3}) is formally
given by $\lim_{p\rightarrow\infty} H_p$. 
\medskip

An alternative formulation is to insert (\ref{5.4}) as an
ansatz into (\ref{5.3}) and separate orders of $\epsilon$. 
This generates the iterative scheme
\begin{eqnarray}
{\cal H}_0 &=& h\nonumber\\
{\cal H}_{p+1} &=& 
-\,\frac{1}{2}\,\sum_{q=0}^{p} 
{\cal H}_q {\cal H}_{p-q} -\,\frac{i}{2}\, 
\dot{\cal H}_p 
\label{5.6}
\end{eqnarray}
for non-negative integer $p$. The corresponding solution $H$ of 
(\ref{5.3}) is given by the series (\ref{5.4}). The first few 
terms are 
\begin{eqnarray}
{\cal H}_1 &=&  -\,\frac{1}{2}\, h^2-\,\frac{i}{2}\,\dot{h} 
\nonumber\\
{\cal H}_2 &=& \frac{1}{2}\, h^3 +\frac{i}{2}\,\{h,\dot{h}\}-\, 
\frac{1}{4}\, \ddot{h} \, , 
\label{5.7} 
\end{eqnarray}
and they clearly fit the structure of (\ref{5.2}). In view of this
scheme one could say that the series (\ref{5.4}) is {\it the 
(unique) solution to (\ref{5.3}) which is analytic in} $\epsilon$. 
\medskip

The convergence of the sequence (\ref{5.5}) implies the
convergence of the series (\ref{5.4}). However, the converse does 
not seem to be necessarily true: it is conceivable that the
limit of the sequence $(H_p)_{p=0}^\infty$ does not exist on account
of effects produced at very large orders in $\epsilon$, while
the operators ${\cal H}_p$ do not ''feel'' these effects and
combine into a convergent series. (A toy-example of such a situation
is given by the scheme $H_0=1$, $H_{p+1}=2\epsilon H_p$, 
where the result is $H_p=(2\epsilon)^p$, which does not
converge as $p\rightarrow\infty$ if $\epsilon=1$, but whose
analogue to (\ref{5.4}) yields ${\cal H}_p=0$, hence $H=0$. We
also note that a modification 
of the scheme (\ref{5.5}) 
by admitting some non-zero starting value $H_0$ 
would change the higher order
behaviour of the $H_p$, while leaving the ${\cal H}_p$ unchanged). 
The actual convergence of 
$\lim_{p\rightarrow\infty} H_p$
would thus imply these higher order effects to 
become successively negligible. The principle possibility
of a situation in which they are not negligible becomes clear
when one tries to define precisely what the dots in (\ref{5.2}) 
mean. All of our considerations will concern statements
given for some arbitrary but finite order in $\epsilon$. 
For this type of statements, both procedures are equivalent, 
but what we mean by ''the'' iterative solution 
$H$ is --- in case of doubt --- the series (\ref{5.4}) rather than
the limit of (\ref{5.5}). 
Nevertheless, the iteration (\ref{5.5}) is more appealing 
as a fundamental issue because it
does not even require the parameter $\epsilon$ to be
different from $1$. In contrast, the procedure 
(\ref{5.6}) relies on a particular
book-keeping of terms which is controlled by $\epsilon$. 
Thus, our hope is that these two procedures are in fact equivalent, 
the parameter $\epsilon$ merely being
a technical tool in computations that could as well be omitted. 
\medskip

In the sense elaborated above, it is clear that the series
(\ref{5.4}) qualifies as a formal solution of the differential
equation (\ref{5.3}) or, after setting $\epsilon=1$,
of (\ref{3.18}). In case of convergence, it will be
an actual solution. 
The same applies to the conditions (\ref{3.19})--(\ref{3.20}). 
The really interesting question, to which we
will now pay our attention, is whether it satisfies 
equation (\ref{4.10}) which guarantees the uniqueness of the
decomposition. In order to investigate this question in a most
economic way, we re-write the equations
(\ref{4.6}) and (\ref{4.10}) by rescaling $h$, $H$ and
$\partial_t$ by a factor $\epsilon$, while leaving
$\delta S$ and $\frac{\delta D}{D}$ unchanged. The variation
of $h$ thus becomes
\begin{equation}
\delta h = [h,\frac{\delta D}{D}] + \,\frac{1}{2}\, \epsilon\, 
[\dot{h},\delta S] + \epsilon\, [h,\delta S]\,\partial_t \, , 
\label{5.8}
\end{equation}
and the uniqueness condition (\ref{4.10}) reads
\begin{equation} 
\delta H = [H,\frac{\delta D}{D}] 
+i\,[H-h,\delta S] +\epsilon\, [H,\delta S]\,(\partial_t + i H)\, . 
\label{5.9}
\end{equation}
We also display the variation of the differential equation (\ref{5.3}), 
\begin{equation}
\delta H = \delta h -\,\frac{1}{2}\,\epsilon\,\{H,\delta H\} 
+\frac{i}{2}\,\epsilon\, [\,[h,\delta S],H] + 
\frac{i}{2}\,\epsilon\, 
[H, \Big(\frac{\delta D}{D}\Big)
\begin{array}{ccc} {\!\!\!\dot{}}\\ {} \end{array}\!\!] 
-\,\frac{i}{2}\,\epsilon\, (\delta H)\,\dot{}\, , 
\label{5.10}
\end{equation}
where the variation of $\partial_t$ has been performed as 
shown in Section 4. We understand that $\delta h$ in 
this equation stands for the expression (\ref{5.8}). 
The variation of the formal iterative solution $H$ may be
computed either directly (by using explicit expressions
like (\ref{5.7})), or by using the iterative scheme 
(\ref{5.6}), 
or by simply iterating (\ref{5.10})
(i.e. successively re-inserting (\ref{5.3}) and (\ref{5.10}) for
$H$ and $\delta H$). All these procedures generate the
coefficients of 
\begin{equation}
\delta H = \sum_{p=0}^\infty \delta {\cal H}_p \, \epsilon^p 
\label{5.11}
\end{equation}
to any desired order. 
The problem under consideration may now be treated using 
(\ref{5.3}) and (\ref{5.8})--(\ref{5.10}), 
without any further reference to the structure of $h$. 
As mentioned above, this is an immediate consequence of how we 
insert powers of $\epsilon$. 
Although we treat $h$ as independent of $\epsilon$, the variation
$\delta h$ involves $\epsilon$. This is just an 
artefact of how we group terms consistently. 
(We recall that the wave equation (\ref{3.9}) is rescaled as 
$i\partial_t \chi=(\frac{1}{2}\epsilon\partial_{tt}+h)\chi$, 
which provides another explanation for the $\epsilon$-dependence 
of $\delta h$). 
\medskip

Once having the above formulae at hand, the argument we need
is surprisingly simple. It is based on the observations that
\par
{\it (i)} inserting $\delta H$ from the uniqueness condition 
(\ref{5.9}) into the variation (\ref{5.10}) of the differential 
equation gives an identity (because (\ref{5.9}) guarantees that 
$H'=H+\delta H$ satisfies its respective differential equation
along with $H$), that 
\par
{\it (ii)} the converse of this, i.e. inserting the variation 
(\ref{5.11}) of the iterative solution
into the uniqueness condition (\ref{5.9}), or, likewise,
inserting and re-inserting $\delta H$ from (\ref{5.10}) into
(\ref{5.9}), 
is just the equation we would like to prove, and that 
\par
{\it (iii)} the iterative structure of our problem implies
that {\it (i)} and {\it (ii)} and essentially identical! 
Note that the statements {\it (i)} and {\it (ii)} hold for 
arbitrary order in $\epsilon$, due to the overall consistency
of the way the various factors $\epsilon$ have been inserted. 
\medskip

Let $\delta H_{(\ref{5.9})}$ denote the rhs of (\ref{5.9}) and
$\delta H_{(\ref{5.10})}$ the rhs of (\ref{5.10}). 
We consider the equation
$\delta H_{(\ref{5.9})}=\delta H_{(\ref{5.10})}$, 
which still contains $H$ and $\delta H$. In all steps we understand
that for $H$ the iterative solution is inserted. Technically,
this is achieved by inserting and re-inserting (\ref{5.3}) until
the desired order in $\epsilon$ is achieved. In the
case of the remaining $\delta H$'s, we have two possibilities: 
Inserting $\delta H =\delta H_{(\ref{5.9})}$ gives the identity 
mentioned in observation {\it (i)}, whereas inserting
$\delta H =\delta H_{(\ref{5.10})}$ iteratively gives the
equation of interest, mentioned in observation {\it (ii)}, that
we want to prove. 
\medskip

We proceed iteratively and begin with $O(\epsilon^0)$. To this order,
the equation $\delta H_{(\ref{5.9})}=\delta H_{(\ref{5.10})}$
reduces to the identity
$[h,\frac{\delta D}{D}] = [h,\frac{\delta D}{D}]$. The
terms containing $\delta H$ are of order $O(\epsilon)$,
hence to not contribute. In other words, no matter which
of the two possibilities for inserting $\delta H$ is chosen, 
we find that $\delta H_{(\ref{5.9})}=\delta H_{(\ref{5.10})}$ is true
to $O(\epsilon^0)$. 
\medskip

Now we consider the equation 
$\delta H_{(\ref{5.9})}=\delta H_{(\ref{5.10})}$ at
$O(\epsilon)$. The remaining $\delta H$'s enter only with their
$O(\epsilon^0)$ part, and at this order both possible choices
give the same result, as we have shown above. Hence, we infer
that $\delta H_{(\ref{5.9})}=\delta H_{(\ref{5.10})}$ is true
to $O(\epsilon)$. 
The following steps proceed exactly along
the same lines. Using all previous orders, we may establish 
the equation
$\delta H_{(\ref{5.9})} = \delta H_{(\ref{5.10})}$ to hold
to arbitrary order 
$O(\epsilon^p)$. The crucial point is that it is not
necessary to specify whether we have to insert 
$\delta H_{(\ref{5.9})}$ or $\delta H_{(\ref{5.10})}$
for the remaining $\delta H$'s, because the equality of both
alternatives has been shown in the foregoing step. 
In other words: The uniqueness condition (\ref{5.9}), with
the iterative solution (\ref{5.4}) inserted, 
precisely yields (\ref{5.11}). We have thus proven the
\medskip

{\bf Theorem: The formal solution (\ref{5.4}) and
its variation (\ref{5.11}) satisfy the uniqueness condition 
(\ref{5.9}) to any finite order in $\epsilon$.}  
\par
The explicit check thereof for the first few orders is
in principle straightforward, but rather time-consuming.
Before the general pattern of the proof has become clear, 
we have explicitly done the computation for $O(\epsilon)$ by
hand and for $O(\epsilon^2)$ and $O(\epsilon^3)$ 
using Mathematica 2.2. 
\medskip

This is our proposal for solving the decomposition
problem. Applying the scheme to concrete models, one will 
have to make sure that the the series (\ref{5.4}) makes sense. 
In Section 7, a number of examples is discussed. 
\medskip

We are now in position to give the first few terms in the
expansion of the quantities ${\cal A}$, ${\cal B}$ and
${\cal K}$ which have proven to determine the unitary 
(Schr{\"o}dinger type) evolution within the spaces
$\H^\pm$ with respect to a WKB-branch. Using the
hermiticity property of $h$ and its time-derivative, it
is easy to compute $H^\dagger$. Then (\ref{3.22}) immediately
yields 
(taking into account the rescaled definition
$\epsilon H+1 ={\cal A}+i{\cal B}$)
\begin{eqnarray}
{\cal A} &= & 1 + \epsilon \,h -\,\frac{1}{2}\,\epsilon^2\,h^2 
+\epsilon^3 \left( 
\frac{1}{2}\,h^3-\,\frac{1}{4}\, 
\ddot{h} \right) + O(\epsilon^4) 
\label{5.12}\\
{\cal B} &=& -\,\frac{1}{2}\,\epsilon^2\,\dot{h} + 
\frac{1}{2}\,\epsilon^3\,\{h,\dot{h}\} +
O(\epsilon^4) \, . 
\label{5.13}
\end{eqnarray}
The first of these equations could provide the key for the 
question whether the positivity property (\ref{3.33}) 
of $\cal A$ that was
necessary to write down the unitary evolution equations 
(\ref{3.38})--(\ref{3.39}) actually holds in concrete models. 
Proceeding formally, 
the operator ${\cal A}^{1/2}$ defining the
effective wave function by (\ref{3.35}) becomes 
\begin{equation}
{\cal A}^{1/2} = 1 + \frac{1}{2}\,\epsilon\, h - 
\,\frac{3}{8}\,\epsilon^2 \,h^2 + \epsilon^3 \left( 
\frac{7}{16}\,h^3 -\,\frac{1}{8}\, 
\ddot{h}\right) + O(\epsilon^4)\, , 
\label{5.14}
\end{equation}
the inverse of which as well as the commutator with $\partial_t$ 
being readily computed. 
Inserting the results into 
into (\ref{3.40}), the evolution operator $\cal K$ 
takes the form 
\begin{equation}
{\cal K}=h -\,\frac{1}{2}\,\epsilon \, h^2 + 
\epsilon^2 \left( 
\frac{1}{2}\,h^3 -\,\frac{i}{8}\, 
[h,\dot{h}] - \,\frac{1}{4}\,\ddot{h} \right) 
+O(\epsilon^3)\, , 
\label{5.15}
\end{equation}
whose hermiticity is evident. 
Whenever the position of $\epsilon$ characterizes quantities
that are actually small (e.g. in a semiclassical context),
these expressions may be used as approximations.
\medskip 

Let us at the end of this Section comment on an alternative
formulation specifying our proposed solution (\ref{5.4}). Recall that 
the differential equation for $H$ is related to the linear 
problem (\ref{3.28}) which now reads 
\begin{equation}
\epsilon^2 \,\ddot{u}+(1+2\epsilon h)u=0\, . 
\label{5.16}
\end{equation}
Any solution thereof
povides a solution to (\ref{5.3}) by 
${\cal P}\equiv 1+ \epsilon H = i\epsilon\dot{u}u^{-1}$. 
It is easy to show that any solution $u$ of the form 
\begin{equation} 
u = v \,e^{-i t/\epsilon}
\quad\qquad{\rm with}\qquad\qquad 
v=\sum_{q=0}^\infty \,v_q \epsilon^q 
\label{5.17} 
\end{equation}
provides the iterative solution (\ref{5.4}) by 
\begin{equation}
H = i \, \dot{v}\,v^{-1} \, . 
\label{5.18}
\end{equation}
Here it is of course 
understood that the coefficient operators $v_q$ do not
depend on $\epsilon$. Although the solution of the form
(\ref{5.17}) is not unique, all possible freedom cancels
in (\ref{5.18}) so as to make $H$ unique. (If $h$ were a 
c-number function of $t$, one could state that (\ref{5.17}) 
is unique up to multiplication by an analyic function of $\epsilon$). 
At the formal level (\ref{5.18}) reproduces the coefficients of the
power series (\ref{5.4}). 
The differential equation satisfied by $v$ turns out to be 
\begin{equation}
i\,\dot{v}=h v + \frac{1}{2}\,\epsilon\,\ddot{v}\, , 
\label{5.19}
\end{equation}
which may be used to compute the coefficients $v_q$ by 
iteratively solving the linear inhomogeneous differential
equations $i\dot{v}_0=h v_0$, 
$i\dot{v}_{q+1}=h v_{q+1}+\frac{1}{2}\ddot{v}_q$ 
(and choosing arbitrary integration constants). 
The formal series emerging from this procedure 
is nothing but the WKB-expansion 
associated with the differential equation (\ref{5.16}) 
to all orders in $\epsilon$ (which serves as the ''small'' 
$\hbar$-type parameter). This analogy becomes even 
more appealing by the fact 
that (\ref{5.16}) formally resembles 
a one-dimensional time-independent Schr{\"o}dinger
equation, with $t$ being the position coodinate 
and $-1-2\epsilon h$ playing the role of the potential. 
In a sense, the computation of $H$
is traced back to a problem within the framework of conventional 
quantum mechanics. 
We may thus be confident that in a large number of concrete models 
our proposed solution $H$ actually exists (i.e. that the
formal power series for $v$ and $H$ belong to analytic 
operator valued functions). 
\medskip

In the literature, the general solution of a differential 
equation of the type $\ddot{u}+Vu=0$ 
is sometimes given as a linear combination 
$v e^{-i t} + w e^{i t}$ or 
$r \sin t + s \cos t$. 
However, the requirement that $u$ is of the form (\ref{5.17}) 
is {\it not} an attempt to say something about the time-dependence 
of $u$ for finite $\epsilon$, neither is it
an asymptotic condition for large $|t|$ mimicking ''positive
frequency'' (we do not even suppose the coordinate $t$ to extend
over all real values). There is no fall-off condition for $v$ 
involved. It may happen, for example, that for some
finite $\epsilon$ the oscillatory factor $e^{-it/\epsilon}$
in $u$ is numerically almost irrelevant as compared to $v$. 
An illustration of this type of solution is provided by 
Example 2 in Section 7 where $H$ is computed for $h=a+b\,t$ 
and $v$ turns out to behave oscillatory and bounded in the domain
$h \gg 0$ but oscillatory and exponentially unbounded 
in the domain $h \ll 0$. 
The nature of $v$ as defined in (\ref{5.17}) without any 
further condition is 
absolutely crucial in our construction:
for generic time-dependence of $h$, one would not
expect a solution of the form (\ref{5.17}) to exist if 
$v$ was supposed to be ''small'' or ''slowly varying''. 
The behaviour of the quantities $u$, $v$ and $H$ should not 
be confused with that of wave function (despite the fact
that (\ref{5.19}) is of the same form as the wave equation 
(\ref{3.9})). 
\medskip

There is another interesting feature related to (\ref{5.17}).
Suppose the time dependence of $h$ is $C_0^\infty$, i.e.
$h$ vanishing identically outside an interval
$t_0<t<t_1$ but being non-zero inside and $C^\infty$ for all $t$. 
As a consequence, $u$ from (\ref{5.17}) 
is purely ''positive frequency'' outside
the interval ($u=c_\pm e^{-i t/\epsilon}$ for $t<t_0$ and $t>t_1$
with $c_\pm$ being independent of $t$). 
On the other hand, (\ref{5.17}) is a differential equation
that can be integated once $u=c_{-}e^{-i t/\epsilon}$ for
$t<t_0$ is known, and the result will in general be such that 
$u\neq c_{+}e^{-i t/\epsilon}$ for $t>t_1$. This tells us that 
there will be no solution $u$ of the required type. 
(This is actually {\it the} argument against the existence of
a preferred standard-type separation into positive and negative 
frequencies in the non-stationary case). 
Since $v$ and 
$H$ may in any case be generated as {\it formal} power series, we
conclude that in the case of $h$ being $C_0^\infty$ in $t$, 
these formal series {\it will not converge} and cannot be associated
with analytic functions in a unique way. Similar effects can be
expected when the basic ingredients of the model
($ds^2$ and $U$) involve $C_0^\infty$ quantities. It is
not quite clear what the correct conditions on the model are 
in order to make sense of the series (\ref{5.4}), but
it is conceivable that they have to do with {\it analyticity}. 
\medskip

\section{Discussion \& Speculations}
\setcounter{equation}{0}

Let us make some speculations concerning
the structure we have possibly touched upon. 
One appealing feature of the iteration procedure defining $H$ 
(preferably in the version (\ref{5.5}), although version
(\ref{5.6}) is technically simpler) is that a (formal) solution 
is specified ''without ever having made a choice''. The procedure
as such just consists in shifting the unwanted $H$-dependence
at the rhs of (\ref{5.3}) to arbitrarily large orders in $\epsilon$.  
The key criterion for solving the problem is the uniqueness
condition (\ref{5.9}). Nevertheless, one would like
to know what is behind such a procedure. 
The whole situation is a bit remniscent of the idea of
general covariance, stating that the geometric objects used to
formulate physical laws can be described in terms of coordinates,
but still have their right own as ''objects''. The different
descriptions (e.g. of a tensor field in various coordinates) 
must match (i.e. obey appropriate transformation laws) so as
to allow for an ''object''-interpretation 
(e.g. as a tensor). 
Considering this as an analogy to our situation, one is tempted
to talk about ''covariance with respect to WKB-branches''. 
The WKB-branch $(S,D)$ thus plays the role of a coordinate
description of a geometric structure. 
The corresponding structure in our case encodes
not only the (unique) decomposition, but also leads to the
unitary evolution equations (\ref{3.38}) and (\ref{3.39}) 
within the two subspaces $\H^\pm$ of wave functions. 
These equations show up only in the context of WKB-branches,
i.e. there is no unique hermitean operator, just as a
scalar field is not given by a unique function of its coordinates 
unless the coordinate system is specified. 
Nevertheless, the structures emerging in WKB-branches,
such as decomposition and unitarity, match 
in a beautiful way. 
\medskip 

We should emphasize here that we have not dealt with
the question how the evolution parameters $t$ relate to
what we experience as time, nor have we suggested a complete
scheme how to reconstruct standard quantum physics and ''the''
Schr{\"o}dinger equation. We merely suggest that, whenever the
decomposition problem may be solved positively
(i.e. the series for $H$ can be given a sense) 
the mathematical structures of standard quantum mechanics
(positive definite scalar product, Hilbert space, unitarity and
the possibility of writing down probability-type expressions) 
begin to come into our reach. It is thus conceivable that
the further steps necessary in deriving quantum mechanics from the
Wheeler-DeWitt equation proceed within this familiar 
mathematical framework. 
\medskip

The most difficult of the questions raised seem to be the 
technical aspects of the convergence properties of the
formal solution (\ref{5.4}). Without trying to go into the details here,
we note that time-derivatives of $h$ of arbitrarily large order are
involved. Thus we expect the solution $H$, whenever it exists,
to rely on the global structure of the model and the
WKB-branch. 
As we have pointed out at the end of the preceding Section
it might be some kind of analyticity condition
that ensures convergence of the series. 
We have seen that at least $C^\infty$ is likely not to be 
the appropriate differentiability class of 
$(ds^2,U)$. 
On the other hand, 
it is conceivable that the need for analyticity is just a technical
tool at some stage in order to give the operator $H$ as provided
by the formal series 
(\ref{5.4}) a definite meaning, and that this condition disappears
when the action of $H$ is maximally extended. This might be
in analogy to the attempt to define the operator 
${\cal U}=\exp(\epsilon\frac{d}{dx})$ by means of its formal power
series. As an intermediate step, one encounters analytic functions, 
but in the end, the action of $\cal U$ may be extended to the whole 
of the Hilbert space $L^2(\R)$, without any reference to analyticity. 
If this analogy holds, the structure responsible for the
preference of a decomposition would be of global (geometric)
nature, rather than relying on analyticity, and
the class of models to which our formalism applies can be expected
to be rather large. 
\medskip

In any case, our proposal seems to transcend the standard
local differential geometric framework (within which we know the 
solution of the generic decomposition problem to be 
negative \cite{Kuchar1}). 
As we will illustrate for the case of a toy-model in the next 
Section (Example 7),  
a naive treatment of the series (\ref{5.2}) or (\ref{5.4}) 
leads to an argument against the existence
of the unique decomposition that might be erraneous 
though intuitively appealing. 
Maybe one could state that, despite the lack of a 
{\it local geometric} structure (such as a symmetry) 
specifying a unique decomposition, some underlying
{\it global} or {\it analytic} structure does the job. 
In the latter case our proposal might give a hint towards a subtle 
role, not yet completely understood, played by analyticity in the 
proper formulation of the laws of nature at the fundamental level. 
\medskip

The major difference between the picture suggested by a 
quantized scalar particle and the picture due to a quantum 
cosmological model has already been addressed above:
In the former, one thinks about the wave function $\psi$ as of
a field propagating causally in space-time, and the
interpretational framework, such as a decomposition of the
space of wave functions,
should be in accoordance with this point of view. 
Thus, whatever
the significance of such a decomposition 
is, it should be determined locally on (or near) some
spacelike 
hypersurface and not depend on what will happen 
to the background $({\cal M},ds^2,U)$ in the far future.
The local character of 
the standard frequency decomposition in generic 
backgrounds refers to such a situation. 
In contrast, the appearance of a formal series in 
our proposal destroys this causal picture. The 
decomposition of $\H$ into $\H^\pm$ is determined by 
global issues in that it cannot be inferred from knowing
$(ds^2,U)$ only near some hypersurface. 
The scalar particle picture can at best lead to locally
defined decompositions, e.g. associated with 
spacelike hypersurfaces. It is conceivable that our proposed
unique decomposition is related to all these
local decompositions by some kind of average. 
(We can even imagine that the actual well-posedness of such
an average is tied to the convergence of
the series for $H$. This would trace back the convergence problem 
to the existence of, say, an appropriate measure). 
Since the idea of causal propagation of the wave function is no 
longer crucial in quantum cosmology, these
considerations are not necessarily objections against our 
approach but might help to uncover the structure which it
is based upon. 
Also, there might be relations to the refined algebraic 
quantization scheme, on which we will speculate in  
Section 8. There we will give a heuristic
argument that this approach seems to {\it require} 
the existence of a preferred decomposition (whose relation to
ours remains open). 
\medskip 

In order to get some feeling for the way how the formal 
solution may become an actual one, 
we consider the simple case in which 
$\dot{h}=0$. In terms of the coordinate system
$(t,\xi^a)$ this means that the operator $h$ does not depend on $t$. 
The flat Klein-Gordon equation 
($ds^2=-d\tau^2+d\vec{x}^2$, $U=m^2$, $S=-m\tau\equiv -t$, $D=1$) 
is a special case,
with $h = -\frac{1}{2m^2}\triangle$. 
(We note that $D=const$ is not a general consequence of
$\dot{h}=0$. In two-dimensional models, 
$n=2$,  one can show that $\dot{h}=0$ implies
$\dot{D}=0$, but even in this case $D$ may depend
on $\xi$). 
For a general situation with $\dot{h}=0$, the iteration
procedure generates a formal solution with $\dot{H}=0$. 
Thus, (\ref{5.3}) reduces to a quadratic equation for $H$.
One may easily check that the series (\ref{5.4}) 
\begin{equation}
H = h -\,\frac{1}{2}\,\epsilon \,h^2 +
\,\frac{1}{2}\,\epsilon^2 \,h^3 - 
\,\frac{5}{8}\,\epsilon^3 h^4 + O(\epsilon^4) 
\label{6.1}
\end{equation}
is just the power series of the closed expression
\begin{equation}
H =\frac{1}{\epsilon}\left( 
\sqrt{1 + 2 \epsilon h}\,-\,1\right) 
\label{6.2}
\end{equation}
around $\epsilon=0$. 
Since one has to insert
$\epsilon=1$, this expression makes sense if 
$h-\frac{1}{2}$ is a non-negative operator, the 
symbol $\sqrt{\,\,\,\,}$ being understood as providing the (unique) 
non-negative square root.  
In the case of the flat Klein-Gordon equation
(\ref{6.2}) leads to the expression $m H=(m^2-\triangle)^{1/2}-m$,
which was already mentioned in Section 2 (and denoted $E$ there). 
It may be well-defined in terms of the Fourier representation,
where $-\triangle$ just becomes $\vec{k}^2$. 
\medskip 

In more general models one will encounter
generalizations of this positivity condition. 
We do not expect these to create major problems, and
as a hint we note that
the condition $h-\frac{1}{2}\ge 0$ is likely to hold 
in realistic models. 
(In the semiclassical context one even has $h^2\ll 1$;
see Ref. \cite{FE8}). 
In general, the structure of (\ref{3.10})--(\ref{3.11}) tells
us that $H^{\rm eff}$ is (at least at the formal level at which 
we are 
treating such issues here) a non-negative operator, and whenever
the ''potential term'' 
$-\frac{1}{2} D(D^{-1})\,\dot{}\,\,\dot{}\,\,$ 
in (\ref{3.10}) is bounded from below, we can expect $h$ to share
this property. 
\medskip

The results obtained for the case $\dot{h}=0$
shed new light on the nature of the 
parameter $\epsilon$ and the way how to re-insert
$\epsilon=1$. 
If $h$ were a number, the series (\ref{6.1}) would converge
only if $|2\epsilon h|<1$. However, the symbol $h$ actually stands 
for all numbers contained in the spectrum of the operator $h$. 
This situation may be illustrated in the flat Klein-Gordon
case $h = -\frac{1}{2m^2}\triangle$. Applying $h$ 
on a function 
\begin{equation}
\chi(\vec{x})=\int dk\, e^{i \vec{k}\vec{x}} \, 
\widetilde{\chi}(\vec{k}) \, ,
\label{6.3}
\end{equation}
it effectively acts as $\frac{1}{2m^2} \vec{k}^2$. Thus, the
closed expression (\ref{6.2}) leads to the appearance of 
$\sqrt{1+\epsilon \vec{k}^2/m^2}$. The power series thereof
converges only if $|\epsilon \vec{k}^2|< m^2$. Thus, we may
apply the series (\ref{6.1}) to a function $\chi(\vec{x})$ 
whose Fourier transform $\widetilde{\chi}(\vec{k} )$ has support
inside a bounded interval $-k_0<k<k_0$. Choosing 
$0<\epsilon<m^2/k_0^2$, the series converges and argees with
(\ref{6.1}). Although $\epsilon=1$ may lie outside the
domain of convergence, it is always possible to analytically
continue the result unambigously along the real line to $\epsilon=1$. 
This procedure may be performed for all functions with
bounded support in momentum space. Since the set of these functions is
dense in any reasonable topology considered for this situation,
the action of (\ref{6.2}) may be reconstructed from the
series (\ref{6.1}) although it will not converge on any
function. 
Thus, it might be the case that in more general situations
the structure of the series (\ref{5.4}) should first be
analyzed for small $\epsilon$, restricted to an
appropriate set of states for which it converges. 
The original value $\epsilon=1$ would then be achieved
by means of analytic continuation.
\medskip

The above considerations seem to impy a suggestion how to proceed
in more general cases. One could define, as a preliminary space, 
the set of all wave functions $\chi$ for which there exists a non-zero
$\epsilon$ (and hence a non-trivial interval of $\epsilon$'s) such that 
\begin{equation}
\sum_{p=0}^\infty \epsilon^p \,({\cal H}_p \chi)
\label{6.4}
\end{equation}
converges in a reasonable sense. The action of the power
series operator (\ref{5.4}) should then be constructed by 
analytic continuation and 
an appropriate extension to (diskrete or continuous) linear 
combinations. 
This is of course only a schematic sketch, and in practice 
various technical problems may occur. 
\medskip

In case the series (\ref{6.4}) for $H$ turns out not to converge
on a sufficiently large number of states (or even on none), 
one might think of applying a regularization scheme. 
In this case one would have to check the branch-independence
of the regularized decomposition. 
One might also speculate on the existence of some mathematical
structure that is poorly accounted for by our power 
series for $H$, with the effect that
the {\it first few terms} give reasonable 
results, whereas the thing {\it as a whole diverges}. 
(Note that a similar issue arises for the perturbation
expansion in ordinary quantum field theory). 
A minimalist point of view would, by the way, consider the
operator $\cal K$ as the most important object for practical 
computations. In case there are models in which its formal
power series (i.e. (\ref{5.15}) to all orders) makes sense, 
whereas (\ref{5.4}) does not, one might be tempted to 
accept non-regularizable infinities in $H$. 
\medskip

Let us close this Section by another speculative remark. 
We can imagine that procedures other than the one
leading to (\ref{5.4}) exist and solve the decomposition problem 
at a formal level. 
(Such procedured could e.g. be based on differently defined
iteration schemes, due to alternative way to group terms by means
of a book-keeping parameter $\epsilon$). 
It is not clear whether different schemes effectively lead to the
same result. In case they do not,  
all of these non-equivalent schemes would provide a setting
within which the ''natural'' construction of a unique decomposition is
possible. They would be regarded as different realizations of what was
called ''covariance with respect to WKB-branches'' above 
(just as different tensor fields might solve some geometric
theory), and should be considered as inequivalent solutions 
to the decomposition problem. 
The naturalness of the construction of a decomposition 
(lacking reference to external structures) may be regarded 
to bear some analogy to, say, the problem 
of finding tensor fields built merely from a metric. 
Since the actual existence of our proposed solution seems to 
rely on further conditions, as discussed above, the number of 
different ''natural'' solutions might depend on the model 
as well. Whether all these may in principle be ''realized'' 
(just as different metrics solving Einstein's field
equations may realize space-time) or whether one of them is
singled out fundamentally, is of course at the moment impossible
to answer for us. 
\medskip

\section{Examples}
\setcounter{equation}{0}

Whenever a WKB-branch $(S,D)$ has been chosen and the operator
$h$ is known, the key procedure in our framework is to evaluate the 
formal solution (\ref{5.4}) of the differential equation (\ref{5.3})
and to determine whether (and on which functions) it 
actually converges. This justifies
the consideration of (\ref{5.3}) in situations in which 
the operator $h$ is simple. The property of $h$ being a second
order differential operator in the coordinates $\xi^a$ may even
be ignored to some extent, and emphasis may be laid on
its dependence on the evolution parameter $t$. 
In the following we list some simple cases that can be treated
exactly, and a scenario illustrating 
a class of models within which our proposal may directly 
be confronted with the standard frequency decomposition 
picture. 
\medskip

{\bf Example 1:} $\dot{h}=0$
\medskip

This it the simplest case, and it has already been studied in 
Section 6. 
The power series (\ref{5.4}) corresponds to the closed
expression (\ref{6.2}). Due to the simplicity of this situation, 
non-commutativity of operators does not play any role here. 
In the case of the flat Klein-Gordon equation $h$ is proportional 
to $ - \triangle$, and $H$ may be given a definite meaning, as
outlined in Section 6. 
\medskip

{\bf Example 2:} $\ddot{h}=0$ and $[h,\dot{h}]=0$
\medskip

This is the special case $h=a+b\,t$, where the operators $a$ and
$b$ do not depend on $t$ and commute with each other. 
We will treat them as numbers, and in particular assume that 
$b$ is invertible and $|b|$ exists.  
The general solution of the linear differential equation
(\ref{5.16}) can be expressed in terms of Airy functions 
(Ref. \cite{AbramowitzStegun}, p. 446). 
Using their asymptotic expansion formulae, the particular solution 
of the form (\ref{5.17}) turns out to be 
(up to an $\epsilon$-dependent factor which is irrelevant) 
\begin{equation}
u = {\rm Bi}(-x) -i \,{\rm sgn}(\dot{h}) \, {\rm Ai}(-x) 
\label{7.1}
\end{equation}
with 
\begin{equation}
x = \frac{1+2\epsilon h}{(2\epsilon^2|\dot{h}|)^{2/3}} \, . 
\label{7.2}
\end{equation}
Hence, the proposed unique solution for $H$ is given by 
\begin{equation}
\epsilon H+1 = \frac{i\epsilon}{u}\,\frac{du}{dt} = 
i\,{\rm sgn}(\dot{h})\, 
\frac{(1+2\epsilon h)^{1/2}}{x^{1/2}\,u}\,\frac{du}{dx}\, . 
\label{7.3}
\end{equation}
This results into the corresponding 
expansion (\ref{5.4}) in terms of $\epsilon$, the various 
${\rm sgn}(\dot{h})$ and $|\dot{h}|$ terms being properly absorbed 
into an expression analytic in $\dot{h}$ (which thus admits
$\dot{h}=0$ as a special case). 
An alternative way to expand $H$ is to keep the combination
$1+2\epsilon h$ and treat only the $\epsilon$ in the denominator
of (\ref{7.2}) as expansion parameter. (This actually amounts to 
re-interpret 
the differential equation (\ref{5.3}) by distinguishing the
two $\epsilon's$ as two different parameters and expanding with
respect to the second one). 
This type of expansion is given by
\begin{equation}
\epsilon H =\left( 
\sqrt{1 + 2 \epsilon h}\,-\,1\right) - 
\,\frac{i}{2}\,\,\frac{\epsilon^2\dot{h}}{1+2\epsilon h} 
+ \frac{5}{8}\,\, 
\frac{(\epsilon^2\dot{h})^2}{(1+2\epsilon h)^{5/2}} 
+O((\epsilon^2\dot{h})^3)\, , 
\label{7.4}
\end{equation}
thus providing a generalization of (\ref{6.2}). The factor
$-\frac{i}{2}$ of the term linear in $\dot{h}$ may be checked
against the second term of ${\cal H}_1$ in the 
general expansion (\ref{5.7}). The existence of this solution
shows that the series (\ref{5.4}), computed by means of the
iteration (\ref{5.6}) with 
$\ddot{h}$ and all higher derivatives neglected and operator
ordering being ignored, is actually the power expansion of an
analytic function. 
\medskip

This example provides an illustration that the form
(\ref{5.17}) is not indented to impose conditions on the
asymptotic behaviour of $v$, because $u$ blows up 
exponentially in the domain $h \ll 0$ while it is
bounded and oscillates for $h \gg 0$. Hence, the
function $v$ is oscillating but has an exponential
prefactor in the domain $\dot{h} t<0$. 
\medskip 

{\bf Example 3:} $h=\,$polynomial in $t$ with commuting 
                 coefficients
\medskip

This case seems to be untractable in general, but 
in principle works like the previous one when again 
use of the linear differential equation (\ref{5.16}) is made. 
The case of $h$ being quadratic in $t$ leads to
parabolic cylinder functions
(Ref. \cite{AbramowitzStegun}, p. 685). In general, 
the power series of the proposed solution $H$ coincides 
with the result of the iteration procedure (\ref{5.6}) 
when all derivatives of $h$ higher than the degree of the
polynimial $P$ are dropped. 
\medskip

{\bf Example 4:} $h=\alpha\, t^{-1}$ 
\medskip 

This case is distinguished by its scaling properties: 
a glance at the series (\ref{5.4}) shows that $H$ is $\epsilon$ 
times a function of the ratio $t/\epsilon$. 
The linear differential equation (\ref{5.16}) leads to a special
case of confluent hypergeometric functions, the solution 
of the form (\ref{5.17}) being (up to
an irrelevant non-integer power of $\epsilon$)
\begin{equation}
u = U(i\alpha,0,z)\, e^{-i t/\epsilon} 
\qquad\qquad {\rm with} 
\qquad\qquad z = \frac{2i}{\epsilon}\, t\, . 
\label{7.5}
\end{equation}
Here, $U$ is the second Kummer function 
(Ref. \cite{AbramowitzStegun}, p. 503)  
\begin{equation}
U(a,0,z)= z^{-a} F(a,z)\qquad
{\rm with}\qquad
F(a,z)=\sum_{q=0}^\infty 
\frac{(a)_q (a+1)_q}{q!}\,\frac{(-)^q}{z^q} \, , 
\label{7.6}
\end{equation}
where  
$(b)_0=1$, $(b)_q=b(b+1)\dots(b+q-1)$ for any $b$. 
The proposed solution of (\ref{5.3}) is given by 
$H = i\dot{u}u^{-1}-\epsilon^{-1}=i \dot{v}v^{-1}$
with $v$ --- as defined in (\ref{5.17}) --- 
being proportional to $U(i\alpha,0,z)$. 
Using the form of $u$ as displayed above, 
we get 
$H=\frac{\alpha}{t} - \frac{2}{\epsilon}F^{-1}\partial_z F$ 
with $z=\frac{2i}{\epsilon}\,t$ inserted. Thus, $H$ is analytic
in $\epsilon$ and in fact coincides with the series
(\ref{5.14}) for $h=\alpha\,t^{-1}$. 
Denoting its functional dependence by 
$H(\alpha,\epsilon,t)$, we have the symmetries 
$H(\alpha,-\epsilon,-t)=-H(\alpha,\epsilon,t)$ and 
$H(\alpha,-\epsilon,t)= - H^*(-\alpha,\epsilon,t)$ 
(all three variables are assumed to be real), hence
$H(-\alpha,\epsilon,-t)=H^*(\alpha,\epsilon,t)$. 
For large $|t|$ the ratio $H/h$ approaches unity. 
\medskip

{\bf Example 5:} $h=\beta\, t^{-2}$ 
\medskip 

This case is particularly important because it emerges in
simple quantum cosmological minisuperspace models. In the 
spatially flat Friedmann-Robertson-Walker (FRW) universe with 
scale factor $a$ (being the only minisuperspace variable) 
and positive cosmological constant $\Lambda$, the DeWitt metric
and the potential are given by 
$ds^2=-da^2$ and $U=a^4 \Lambda$. 
The WKB-branch is specified by 
$S=-\frac{1}{3}a^3\sqrt{\Lambda}=-t$ and 
$D= t^{-1/3}$ (it is unique up to $S\rightarrow \pm S+const$ and
$D\rightarrow const\, D$),
the range of the ''time'' coordinate being $t>0$. 
Equation (\ref{3.10}) gives (setting $H^{\rm eff}=0$) 
$h= \frac{1}{9} \,t^{-2}$. 

A slightly more sophisticated model is obtained by putting a
massless scalar field $\phi$ into the FRW universe, thus
$ds^2=-da^2+a^2 d\phi^2$ and $U=a^4 \Lambda$. The action is 
chosen as before 
$S=-\frac{1}{3}a^3\sqrt{\Lambda}=-t$, 
whereas $D=t^{-1/2}$. 
The variable $\phi$ plays the role of the coordinates
$\xi^a$, and applying 
(\ref{3.10})--(\ref{3.11}) one easily finds 
$h= \frac{1}{2} \,t^{-2}(-\frac{1}{9}\, \partial_{\phi\phi}+ 
\frac{1}{4})$. For any $t>0$ this is a positive operator on
the hypersurface $\Sigma_t$ (i.e. acting on functions $\chi(\phi)$). 
In more realistic models it is often extremely difficult to
specify an exact solution of the Hamilton-Jacobi 
equation, but we can expect a $t^{-2}$ behaviour in $h$ 
to play a dominant role as well. 
\medskip 

The general solution of the linear problem (\ref{5.16}) 
for $h=\beta\, t^{-2}$ may be 
expressed in terms of Bessel functions
(Ref. \cite{AbramowitzStegun}, p. 358). 
Confining ourselves to the range $t>0$ for the
moment, the combination 
involving the second Hankel function 
\begin{equation}
u = \sqrt{z} \,H^{(2)}_\nu(z)\equiv 
\sqrt{z}\left( J_\nu(z)- i Y_\nu(z)\right)
\label{7.7}
\end{equation}
with 
\begin{equation}
z=\frac{t}{\epsilon}
\qquad\qquad  {\rm and} 
\qquad\qquad 
\nu^2 = \frac{1}{4} - \frac{2\beta}{\epsilon} 
\label{7.8}
\end{equation} 
is of the form (\ref{5.17}),  
up to an irrevevant $\epsilon$-dependent factor. 
Expanding $u$ for small positive $\epsilon$ at constant $\nu$ 
and positive $t$ (i.e. using the asymptotic expansion 
of Bessel functions for $z\rightarrow \infty$) 
gives an expression of the structure
$c v e^{-it/\epsilon}$ where $c$ depends on $\nu$ and $\epsilon$ 
but not on $t$ (and is thus irrelevant for our purposes) and
$v$ is a power series in $\epsilon$ that depends on $\nu$ only 
through $\nu^2$. Inserting the second equation of (\ref{7.8})
into $v$ yields a power series in $\epsilon$. Note that
$\beta$ may be of either sign or zero, and $\nu$ will be
real or imaginary, depending on the values of $\beta$ and $\epsilon$.  
The proposed solution for $H$ is given by 
$H = i\dot{u}u^{-1}-\epsilon^{-1}$ or, likewise, by 
(\ref{5.18}), its expansion in powers of $\epsilon$ reproducing
the formal solution (\ref{5.4}) for $h=\beta\, t^{-2}$. 
In the closed expression for $H$, any of the two 
solutions 
for $\nu$ may be inserted because the ratio 
$H^{(2)}_\nu(z)^{-1}\partial_z H^{(2)}_\nu(z)$ 
is invariant under the substitution $\nu\rightarrow -\nu$. 
For fixed positive $\epsilon$, the ratio $H/h$ approaches 
unity in the limit $t\rightarrow\infty$. 
Due to singluarities in the Bessel functions, the closed expression
is convenient only for positive $t$ and $\epsilon$. Nevertheless,
the power series reveals that $H$ is analytic in $\beta$,
$\epsilon$ and $t^{-1}$. 
Denoting this dependence by $H(\beta,\epsilon,t)$, the 
other ranges of variables are provided by the 
identities 
$H(\beta,\epsilon,-t) = H^*(\beta,\epsilon,t)$ and
$H(\beta,-\epsilon,t) = - H^*(-\beta,\epsilon,t)$ 
where $\beta$, $\epsilon$ and $t$ are understood to be real. 
The simultaneous change of sign of these three variables changes
the sign of $H$. 
\medskip 

{\bf Example 6:} $h=k+\beta\, t^{-2}$ 
\medskip 

This case may be treated in complete analogy to the previous one. 
The relevant solution of (\ref{5.16}) is given by 
\begin{equation}
u = \sqrt{t} \,H^{(2)}_\nu(c t) 
\qquad {\rm with}
\qquad c=\frac{1}{\epsilon}\sqrt{1+2\epsilon k} 
\label{7.9}
\end{equation}
and $\nu$ as before. The preferred decomposition is
therefore defined by 
$H = i\dot{u}u^{-1}-\epsilon^{-1}$. 
The factor $c$ replacing
$\epsilon^{-1}$ in the argument of the
Hankel function has the effect already encounted in the
first example that the ratio $H/h$ does not aproach unity
but $(\epsilon k)^{-1}(\sqrt{1+2\epsilon k}-1)$ 
for large $t$. 

A further possible generalization is
$h=k+\alpha\,t^{-1}+\beta\,t^{-2}$ which leads to Whittaker 
functions (Ref. \cite{AbramowitzStegun}, p. 505). 
\medskip

{\bf Example 7:} $h$ being asymptotically constant in two regions 
\medskip

Here we would like to discuss an 
argument that might be raised against the existence of a
preferred decomposition. 
Suppose $h$ is a c-number function of $t$ 
converging to a constant $h_\infty$ as 
$|t|\rightarrow\infty$ but undergoing some
non-trivial smooth change at finite times. 
For simplicity we assume $h_\infty=0$ and set $D=1$. 
Naively, one would expect the form of the series (\ref{5.4}) 
to indicate that $H$ vanishes asymptotivally for large $|t|$. 
If this were true, the proposed
decomposition could be thought of agreeing 
with the standard asymptotic 
flat space-type decomposition 
into positive and negative frequencies 
($\psi^\pm\sim e^{\mp i t}$)
in the ''far past'' region 
$t\rightarrow -\infty$ {\it and at the same time} it 
would agree with the according frequency 
decomposition 
(again $\psi^\pm\sim e^{\mp i t}$)
in the ''far future'' region 
$t\rightarrow\infty$. However, we know that in case of
a nontrivial time-dependence of $h$ in some intermediate
region, these two frequency decompositions are different.
This is because a wave function starting with, say, 
positive frequency in the far past ($\psi\sim e^{-it}$)
will ''feel'' the time-dependence of $h$ and in general
evolve into a linear combination of both frequencies $e^{\mp i t}$ 
at late times. (In terms of the physics of a scalar field, 
this indicates the production of particles
by the time-dependent background. In terms of mathematics, 
it is represented by a Bogoljubov transformation). 
Thus, the argument states that the proposed preferred 
decomposition 
coincides with the two asymptotic ones which are however 
different: this is obviously a contradiction. 
The answer is presumably that the expectation 
$H\rightarrow 0$ in the asymptotic regions
is in general not justified. 
In the Examples 4, 5 and 6 treated above, $h$ vanishes for
large $|t|$ but has a singularity at $t=0$, so that 
these cases are no good illustrations for the issue
under consideration here. 
A better example would be $h=\beta (1+t^2/a^2)^{-1}$. 
In this case either $H$ does not exist 
(its formal series (\ref{5.4}) diverging for all $\epsilon\neq 0$
and for all $t$) 
or does not
vanish asymptotically. In any case, the non-local character of
the principle specifying $H$ is clear. 
Although having no explicit example at hand, 
these considerations should illustrate how our proposal 
might be capable to transcent the point of view that a preferred 
decomposition necessarily relies on local geometric structures. 
In view of the speculations made in Section 6, the 
preferred decomposition --- in case it exists --- 
might turn out to be an average 
over the asymptotic frequency decompositions.  
\medskip

\section{Relation to refined algebraic quantization?}
\setcounter{equation}{0}

The results achieved in this paper have been of a 
rather formal nature, and the precise condition under which
they materialize into well-defined mathematical objects
is not known (apart from the speculation that it refers to
the global structure of minisuperspace and possibly has 
something to do with analyticity). This makes it difficult
to re-express our findings in terms of an underlying 
structure or object for which the equations we wrote
down are just representations with respect to local 
WKB-branches. 
\medskip

Nevertheless, in pursuing the subject, one might gain some
unified view of what happens when different quantization and
interpretation schemes are applied. As already mentioned in 
Section 2, the interpretational framework associated
with the wave equation (\ref{2.2}) can be based either on 
the $L^2$-scalar product $q$ from (\ref{2.7}) or on the 
Klein-Gordon type scalar product $Q$ from (\ref{2.8}), 
the latter being particularly appropriate for the 
scalar particle quantization picture. Our approach is in some
sense a mixture between the use of $Q$ (which admits the
notion of causal propagation on the manifold $\cal M$) and
a proposal for solving the decomposition problem that is motivated
by quantum cosmology and transcends the notion of causality. 
\medskip

This opens perspectives for a deeper understanding of the
relation between various structures associated with the
wave equation (\ref{2.2}). With regards to our proposal for
fixing a decomposition $\H=\H^+\oplus\H^-$, the most natural 
question seems to be whether the space
$(\H,Q_{{}_{\rm phys}})\equiv(\H^+,Q)\oplus (\H^-,-Q)$ is closely
related (or even identical to) the Hilbert space of physical states
$({\cal H}_{{}_{\rm phys}},\langle\,|\,\rangle_{{}_{\rm phys}})$  
as constructed by the refined algebraic quantization 
approach, i.e. to what extent the results achieved for the
flat Klein-Gordon equation 
\cite{Aetal,Higuchi2,Marolf1}, 
as mentioned in Section 2, 
carry over to more general models. In the positive case, 
our approach seems to provide a piece of information that does not
show up explicitly in the refined algebraic scheme, 
namely the decomposition, 
i.e. the natural appearance of two orthogonal subspaces $\H^\pm$ 
spanning the whole of $\H$. 
Even if the answer to this question is more complicated, one
can reasonably expect to learn more about the relation between 
$Q$, $q$ and $\langle\,|\,\rangle_{{}_{\rm phys}}$. 
Eventually, one might be able to reconcile the different approaches
in terms of a single, unified understanding of quantization. 
\medskip

Although we have not entered into the formalism of the 
refined algebraic quantization, there is a heuristic argument
relating the inner (positive definite) product 
$\langle\,|\,\rangle_{{}_{\rm phys}}$ 
to the (indefinite) Klein-Gordon type scalar product $Q$. 
(A construction of the type we will now describe seems to have been 
written down first in the contex of de Sitter space by Nachtmann 
long ago \cite{Nachtmann}. In a more general context  
a similar reasoning appears 
in the work of Rumpf und Urbantke \cite{RumpfUrbantke}). 
Although the Hilbert space 
${\cal H}_{{}_{\rm phys}}$ of
physical states emerging in this 
quantization scheme is constructed in a rather abstract way, 
is seems likely \cite{Donprivate} 
that it can be represented as a set of 
solutions to the wave equation (\ref{2.2}) --- modulo 
the issue of completion ---, i.e. that it
essentially agrees with $\H$.  
The space $\H$ thus carries two
scalar products, $\langle\,|\,\rangle_{{}_{\rm phys}}$ and $Q$. 
We may find some basis $\{\Xi_r\}$ of wave functions which are
othonormal with respect to the former scalar product, 
i.e. $\langle \Xi_r|\Xi_s\rangle_{{}_{\rm phys}}=\delta_{rs}\,$. 
Generic states may be expanded as $\psi=\sum_r \psi_r \Xi_r$. 
The matrix $K_{rs}=Q(\Xi_r,\Xi_s)$ defines a
linear operator $K$ by $(K\psi)_r =\sum_s K_{rs}\psi_s$ 
which relates the two scalar products by 
\begin{equation}
Q(\psi, \phi) = 
\langle\psi| K \phi\rangle_{{}_{\rm phys}} \, . 
\label{8.1}
\end{equation}
The properties of $Q$ imply that the matrix $K_{rs}$ (and
hence the operator $K$) is hermitean with respect
to $\langle\,|\,\rangle_{{}_{\rm phys}}$ and invertible. 
(Actually $K$ is invertible if $Q$ is non-degenerate,
i.e. if $Q(\psi,\phi)=0$ for all $\psi$ implies $\phi=0$. 
There are situations $(ds^2,U)$ in which this is not true, i.e.
in which harmless initial data  
$(\psi|_\Sigma,n^\alpha\nabla_\alpha\psi|_\Sigma)$ 
may lead to an exponential blow-up with time and thus have to be 
excluded from $\H$. Accordingly, $Q(\psi,\phi)=0$ for
all $\psi\in \H$ with $\phi\neq 0$ becomes possible 
\cite{Rumpf1}. 
Athough this is not likely to happen in reasonable quantum 
cosmological models, it restricts the number of mathematically
admissible models for which our argument applies. Hence, we
assume the model under consideration to be such that $Q$ is
non-degenerate).
If an operator $K$ defined by (\ref{8.1}), it is unique 
because $\langle\psi| K \phi\rangle_{{}_{\rm phys}}$ is known
for all $\psi$ and $\phi$, and its matrix elements with
respect to the basis $\{\Xi_r\}$ are just $K_{rs}$. 
Since $K$ is invertible, its spectrum 
decays into a positive and a negative part. (Thereby we actually 
assume that $K$ is self-adjoint with respect to
$\langle\,|\,\rangle_{{}_{\rm phys}}$, a point that should be made more 
rigorous in a deeper investigation). The space $\H$ may thus
be expected to decompose into two subspaces $\H^\pm$, associated
with the positive and negative part of the spectrum, respectively. 
In other words, $\H^\pm$ is
spanned by the (generalized) eigenvectors of $K$ to 
positive/negative (generalized) eigenvalues 
(the latter appearing in pairs $(\lambda,-\lambda)$). 
In addition, there are good chances for the spectrum of $K$ 
to be bounded off zero. (For the flat Klein-Gordon equation 
only {\it one} pair $(\lambda,-\lambda)$ with
$\lambda\neq 0$ occurs). 
The two supspaces $\H^\pm$ 
are orthogonal to each other with respect to 
$\langle\,|\,\rangle_{{}_{\rm phys}}$ and hence, using (\ref{8.1}), 
also with respect to $Q$. Moreover, it is clear that
if this construction can be made rigorous, complex conjugation
maps $\H^+$ onto $\H^-$ and {\it vice versa} in a bijective way. 
This argument shows that the inner product 
$\langle\,|\,\rangle_{{}_{\rm phys}}$ is likely to serve
as a structure {\it in addition to the local geometry of
minisuperspace} and to single out a preferred decomposition. 
Since the procedure leading to 
$\langle\,|\,\rangle_{{}_{\rm phys}}$ in the refined algebraic
quantization scheme is essentially unique, the decomposition
derived from (\ref{8.1}) should be unique as well. Hence, 
it should rely on the ''geometry'' of
minisuperspace in some {\it global} sense. 
It is however unclear whether this decomposition coincides 
with the one proposed in Section 5, i.e. the one based on the
formal solution (\ref{5.4}) for $H$. 
\medskip

This point --- at least as a theoretical possibility ---
seems to have escaped the attention of several authors that have
contributed to the development of the refined algebraic quantization
approach. For example, Higuchi \cite{Higuchi2} quotes 
Kucha$\check{\rm r}$'s discussion \cite{Kuchar1,Kuchar2} 
that the absence of symmetries of the Wheeler-DeWitt
equation does not allow a unique separation of positive- and
negative-frequency solutions and simply states that his method
(which is essentially the refined algebraic approach) 
{\it ''does not require such a separation''}. 
Also, Ashtekar {\it et al} \cite{Aetal} 
remark that the construction of ${\cal H}_{{}_{\rm phys}}$ (in the 
context of the Klein-Gordon equation on a curved background
space-time) 
{\it ''may come as a surprise to some readers, as it
seems to violate the accepted idea that there is no well-defined 
notion of a single particle in a non-stationary space-time''}.
Their resolution of this paradox is that 
the new quantum theory {\it ''contains no notion of a conserved 
probability
associated with Cauchy surfaces, as our particle appears to 
'scatter backwards in time' when it encounters a lump of
space-time curvature''}. 
However, when the heuristic argument given above holds, 
this situation is actually reversed: 
The mere co-existence of $Q$ and 
$\langle\,|\,\rangle_{{}_{\rm phys}}$ implies the existence
of a unique preferred decomposition, 
and there {\it is} a well-defined notion of a conserved
(not necessarily positive) probability-type density, provided by the
scalar product $Q$ (namely the integrand of (\ref{2.8})). 
What {\it is} missing in this framework is of
course the notion of causality underlying the decomposition. 
Maybe one could say that the group average applied in this 
quantization procedure has singled out a decomposition 
by some kind of average of global type. 
\medskip

In the case of the flat Klein-Gordon equation we actually know 
the above argument to hold, and in addition we know that $K^2$ is a 
multiple of 
the identity operator ($\langle\,|\,\rangle_{{}_{\rm phys}}$ therefore 
being normalized such that $K^2=1$). This implies that 
$\langle\,|\,\rangle_{{}_{\rm phys}}$, when restricted to 
$\H^\pm$, is just $\pm Q$. We do not know whether this feature
carries over to more general cases. In general, we can only give
a symbolic expression for $K$. 
As already mentioned in Section 2, physical 
states are represented as 
$\psi_{{}_{\rm phys}}=\delta(C)\psi$ with $C$ the wave operator 
in (\ref{2.2}) and $\psi$ a more or less arbitrary function
(not satisfying the wave equation). Defining an auxiliary
operator $\widetilde{K}$ by 
$(K\psi)_{{}_{\rm phys}}=\delta(C)\widetilde{K}\psi$, one finds
\begin{equation}
\widetilde{K} = -\,\frac{i}{2}\,
\{ \delta_\Sigma,n^\alpha \nabla_\alpha \} \,
\delta(C) \, ,
\label{8.2}
\end{equation}
where $\Sigma$ is a hypersurface on which $Q$ is computed, 
$\delta_\Sigma(y)$ the $\delta$-function on $\Sigma$, such that
$\int_{\cal M} d\mu\,\delta_\Sigma \,\varphi=\int_\Sigma d\Sigma 
\,\varphi$ for 
any (test)function $\varphi$ on $\cal M$, and $d\Sigma$ the 
scalar hypersurface element on $\Sigma$. The unit normal to
$\Sigma$ is $n^\alpha$, so that 
$d\Sigma^\alpha=d\Sigma\,n^\alpha$ is the hypersurface 
element as used in (\ref{2.8}). The expression (\ref{8.2}) 
--- when treated rigorously --- should be
independent of $\Sigma$. In flat space, its Fourier
transform may be written down explicitly, exhibiting the
expression $\delta(k^\mu k_\mu+m^2)$. This is identical to
$\frac{1}{2} \,k_0^{-1}(\delta(k_0-\omega_{\vec{k}}) - 
\delta(k_0+\omega_{\vec{k}}))$
with $\omega_{\vec{k}}=(\vec{k}^2+m^2)^{1/2}$, which illustrates 
the origin 
of the change of sign in the negative frequency sector. 
\medskip

As stated above, it is unclear how our heuristic 
argument leading to $K$ relates to the
formalism poposed in the main part of this paper, and how
an {\it analyticity} condition might appear. 
The main condition for the refined algebraic quantization
scheme to work is the self-adjointness of the wave operator
$C$ with respect to the scalar product (\ref{2.7}), 
and this is essentially a condition on the metric and 
the potential, without
any obvious relation to analyticity issues. 
\medskip 

\section{Outlook}
\setcounter{equation}{0}

There is another important issue that seems worth being 
pursued. In this paper we have confined ourselves to the
case that the potential $U$ is positive everywhere in 
minisuperspace. As a consequence, the classical trajectories
(whose tangent vector squared has the opposite sign as $U$)
were always timelike, and we could clearly distinguish 
between incoming and
outgoing modes at the classical level. 
Technically, we have used $\sqrt{U}$ in various 
expressions.
However, in most realistic models the potential may take both signs. 
It is thus desirable to know how far the findings of this paper
may be generalized for these models. 
One might still have a classically well-defined notion of
outgoing and incoming traejctories (as long as $\cal M$ is
foliated by spacelike hypersurfaces), but the classical trajectories
would be spacelike in the $U<0$ domains (which are sometimes
called ''Euclidean''). Thus, a classical trajectory could start with
outgoing orientation, dive into a Euclidean domain and re-enter
the $U>0$ domain with incoming orientation. This might be a hint
for the breakdown of the framework developed in this paper. 
In particular, evolution equations like (\ref{3.23}) would
not easily be defined, and one might have to invoke the 
mathematics of signature changing metrics. Also, 
one could try to save the notion of decomposition by making it
a local concept. On the other hand, the hints at analyticity
playing a deep role in our framework open another possibility. 
Since analytic functions exist ''as a whole'', one could think of 
applying our formalism to the $U>0$ region and look whether it
naturally ''extends'' to the whole of $\cal M$ (maybe in 
some complexified sense). Note that, for example, by describing 
a hypersurface $\Sigma_t$ in terms of some (real) analytic 
function in the domain 
$U>0$, a (complex) global behaviour is fixed at the same time. 
This might help clarifying the role of Euclidean domains and
the complexified concepts (like Euclidean trajectories and
Euclidean action) that are sometimes used in this field. 
\medskip

As a last issue we mention that the most ambitious goal would 
of course be to carry over our proposal to the framework of the 
Wheeler-DeWitt equation in full (rather than mini-) superspace. 
In the metric representation
\cite{Wheeler+DeWitt}, one would have to deal with functional 
differential equations, and the analogues of evolution equations like
(\ref{3.23}) and (\ref{3.38}) would be of the Schwinger-Tomonaga 
rather than the Schr{\"o}dinger type, i.e. involve the feature of 
many-fingered time. Also, regularization issues are likely do
make this a very hard job. As an alternative, one could 
try to work in the connection representation 
\cite{Ashtekar} 
and ask whether the notions of Klein-Gordon type scalar
product and decomposition can be given any sense therein. 
\\
\\
\\
{\Large {\bf Acknowledgments}}
\medskip

I would like to thank Donald Marolf and 
Jos$\acute{\rm e}$ Mour$\tilde{\rm a}$o for stimulating my 
interest in the refined algebraic quantization program during 
the early stages of this work. 
Helmut Rumpf pointed out to me that in certain cases the
operator $K$ defined in Section 8 has $0$ in its point 
spectrum \cite{Rumpf1}. 
Also, I am indebted to Helmut Rumpf and Helmuth Urbantke for bringing 
the predecessors of the refined algebraic quantization program 
\cite{Nachtmann}--\cite{Rumpf1} to my attention. 

\end{document}